\documentclass[prd,twocolumn,showpacs,preprintnumbers,amsmath,amssymb]{revtex4}

\usepackage{graphicx}
\usepackage{dcolumn}
\usepackage{bm}
\usepackage{longtable}
\usepackage{afterpage}

\def\lsim{\mathrel{\lower2.5pt\vbox{\lineskip=0pt\baselineskip=0pt
                      \hbox{$<$}\hbox{$\sim$}}}}
\def\gsim{\mathrel{\lower2.5pt\vbox{\lineskip=0pt\baselineskip=0pt
\hbox{$>$}\hbox{$\sim$}}}}

\newcommand{\ima}{{\mbox{Im}\,}}

\newcommand{\gev}{{\mbox{GeV}\,}}

\newcommand{\be}{\begin{equation}}
\newcommand{\ee}{\end{equation}}

\newcommand{\NP}[1]{Nucl.\ Phys.\ {#1}}

\newcommand{\PR}[1]{Phys.\ Rev.\ {#1}}

\begin{document}

\title{Strange and non-strange quark mass dependence of elastic light resonances 
from SU(3) Unitarized Chiral Perturbation Theory to one loop}

\author{J. Nebreda and J.R. Pel\'aez\\
\emph{Departamento de F\'{\i}sica Te\'orica II. Universidad Complutense. 28040 Madrid. Spain}
}

\begin{abstract}
We study the light quark mass dependence 
of the $f_0(600)$, $\kappa(800)$, $\rho(770)$ and
$K^*(892)$ resonance parameters generated from
elastic meson-meson scattering using unitarized one-loop 
Chiral Perturbation Theory. 
First, we show that it is possible to fit simultaneously all experimental
scattering data up to 0.8-1 GeV together with 
lattice  results on decay constants and scattering lengths up to a pion mass of
400 MeV, using chiral parameters compatible with existing determinations. 
Then, the strange and non-strange quark masses are varied 
from the chiral limit up to values of interest for
lattice studies. In these amplitudes,  the mass and width of the $\rho(770)$ and
$K^*(892)$ present
a similar and smooth quark mass dependence.
In contrast, both
scalars present a similar non-analyticity at high quark masses. Nevertheless,
the $f_0(600)$ dependence on the non-strange quark mass
is stronger
 than for the $\kappa(800)$ and the vectors.
We also confirm the lattice assumption of quark mass independence
of the vector two-meson coupling that, in contrast,
is violated for scalars.  As a consequence, vector widths are very well approximated
by the KSRF relation, and their masses are 
shown to scale like their corresponding meson decay constants.
\end{abstract}

\maketitle


\section{Introduction}

Although QCD is well established as the theory of strong interactions, 
the fact that its coupling becomes large at energies below 1-2 GeV
keeps the hadronic realm beyond the reach of perturbative calculations.
In that regime, lattice methods are a useful tool to
calculate QCD observables, although the discretization involved in this
technique introduces complications of its own, in particular related to
chiral symmetry breaking and the implementation of realistic
small masses for the light quarks.  
Despite the remarkable success of lattice studies, results on light  meson
resonances are few and usually 
obtained at very large quark masses compared with their
physical values \cite{Aoki:1999ff,scalars}. This is 
particularly so for the light scalars, very relevant for nuclear attraction,
but  whose calculations are hindered 
by the so called ``disconnected diagrams''.
Very recently \cite{Hanhart:2008mx}, 
an alternative technique, based on 
Chiral Perturbation Theory (ChPT) and dispersion relations,
has been applied to calculate the 
dependence of the $f_0(600)$ (or ``sigma'') and
$\rho(770)$ resonances on the pion mass 
-- in practice, the average mass of the $u$ and $d$ quarks. 
Now the starting parameters are physical and resonances 
appear in amplitudes that describe real data on $\pi\pi$ scattering. 
The predicted dependence for the $\rho(770)$ compares remarkably
well with previous and later lattice predictions. For the scalar sigma
it shows a non-analyticity that should be taken into account
when extrapolating future lattice data to physical values. 
In this work we extend this study to include the strange quark mass
within an SU(3) ChPT formalism. Our aim is threefold: first,
to confirm previous results within a more general formalism.
Second, to analyze the  dependence on the average mass of the 
 $u$ and $d$ quarks of the $K^*(892)$ and
$\kappa(800)$ strange resonances. The latter, despite being a scalar,
and very similar to the $f_0(600)$,
is much more feasible for lattice calculations
within the next few years \cite{Nagata:2008wk} due to its nonzero
isospin and strangeness. Third,
we also study the dependence of all the
$f_0(600)$, $\kappa(800)$, $\rho(770)$ and
$K^*(892)$ parameters in terms of the strange quark mass.
Finally, let us remark that the dependence of hadronic
observables, meson masses in particular, is not only
of relevance for lattice calculations, 
but also for anthropic considerations \cite{AP} or
the study of the cosmological variability of fundamental constants 
\cite{Cosmology}.

Thus, in the next two subsections we 
introduce very briefly the basic notation of ChPT,
explain the relation between pseudoscalar meson and quark masses, and
review the unitarization procedure. In section II, we
show the fits to the existing experimental data on elastic
scattering as well as to lattice results
 on pion and kaon masses, their decay constants, and 
scattering lengths on the highest isospin channels.
Section III is devoted to the dependence of light 
resonance properties on the non-strange quark masses. 
In section IV we then study the dependence with the strange quark mass
and in Section V we present our summary and conclusions.

\subsection{Chiral Perturbation Theory}
\label{sec:ChPT}

As is well known, pions, kaons and etas can be identified
with the Nambu-Goldstone bosons (NGB) of the spontaneous 
chiral symmetry breaking of QCD. If quarks were massless, so should be the
NGB and they would be separated by a mass gap of the order of 1 GeV from
other hadrons, thus becoming the only relevant QCD degrees of freedom 
at low energies. Of course, quarks are not massless, but the 
$u, d$ and $s$ flavors have a sufficiently light mass to be considered as a perturbation. It is thus possible to write a Low Energy Effective Lagrangian
out of pion, kaon and eta fields,
known as Chiral Perturbation Theory (ChPT)\cite{Gasser:1984gg}. This Lagrangian
is built as the most general derivative and mass expansion that respects 
the symmetries of QCD, particularly its chiral symmetry breaking pattern. 
Except for the leading order, fixed by symmetry and 
the scale of spontaneous symmetry breaking, all terms in the Lagrangian
are multiplied by a low energy constant (LEC) that contains the information on the underlying QCD dynamics and also renormalizes the loop diagrams 
with vertices from lower orders.
In this way, pion, kaon and eta observables 
are obtained as a {\it model independent} expansion in powers
of momenta and masses over the chiral scale $4\pi f_0\simeq 1.2\,$GeV,
where $f_0$ is the pion decay constant in the chiral limit (as it is customary,
for quantities at leading order in the quark mass expansion
 we will use the 0 subscript).

In particular, partial wave amplitudes for 
elastic meson-meson scattering are obtained within ChPT
as an expansion
\begin{equation}
t(s)=t_2(s)+t_4(s)+.... , \qquad t_{2k}=O(p^{2k}),
\label{eq:ChpTseries}
\end{equation}
where $p$ denotes either momenta or meson masses.
Actually, these partial waves carry definite isospin $I$ and 
total angular momentum $J$, but we have momentarily suppressed
these labels for clarity.
As we have just commented,  the leading order $t_2(s)$ corresponds
to the current algebra results and only depends on the scale
of spontaneous chiral symmetry breaking $f_0$.
The next to leading order $t_4(s)$ contains one-loop diagrams
made of vertices from the lowest order Lagrangian, plus tree level diagrams
of $O(p^4)$. Within the SU(3) formalism,
these tree level diagrams are multiplied by 
LECs, denoted as $L_i$, which are independent of 
masses or momenta, and have been determined
from different experiments. In Table~\ref{tab:LECs} we provide 
several sets for the eight $L_i$ that appear in meson-meson scattering
to one loop. Those with an ``r'' superscript carry a 
dependence on the regularization scale $\mu$ \cite{Gasser:1984gg},
customarily chosen at $\mu=M_\rho$. Of course, that
scale dependence cancels 
in the calculation of physical observables.
The values in the second column come from the 
``Main Fit'' of a $K_{l4}$ analysis
to two loops \cite{Amoros:2001cp}, whereas those in the third column come from the same 
reference, but to one loop. Naively one would expect 
the LECs obtained in our unitarized one-loop fits
to lie somewhere in between these two sets of values, since unitarization
reproduces one of the most relevant numerical contributions
from the two-loop calculation, namely the s-channel leading logs.
As one of our main interests is $\pi K$ scattering and 
the $K^*(892)$ and $\kappa(800)$ resonances, we also provide the values obtained from
a very rigorous treatment of $K\pi$ scattering lengths in terms 
the Roy-Steiner dispersion relations \cite{Buettiker:2003pp}.
The rest of the columns correspond to unitarized ChPT fits
that we will explain in Sect.\ref{sec:fits}

\begin{widetext}
\setlength{\LTcapwidth}{5in}
\begin{longtable}{ccccccc}
\caption{\label{tab:LECs}$O(p^4)$ chiral parameters ($\times10^{3}$)
evaluated at $\mu=M_\rho$.
The second and third columns come from the two and one loop analysis listed in \cite{Amoros:2001cp}, where $L_4$ and $L_6$
were set equal to zero.
The fourth column comes from a careful $\pi K$ dispersive analysis \cite{Buettiker:2003pp} using the Roy-Steiner formalism.
The IAMIII column is one of the sets obtained from an older fit with  
the coupled channel IAM \cite{Pelaez:2004xp} (only statistical uncertainties are shown). The columns labeled Fit I and Fit II correspond to the simultaneous fit to experiment and lattice data performed in this work, 
which are described in Sec.\ref{sec:fits} together with their uncertainties.
}\\
    \hline \hline
LECs & Ref.\cite{Amoros:2001cp} $O(p^6) $ & Ref.\cite{Amoros:2001cp} $O(p^4) $ & Ref.\cite{Buettiker:2003pp} & IAM III &\ \ Fit I\ \ &\ \ Fit II\ \ \\
\hline
$L_1^r$ & 0.53 $\pm$ 0.25 & 0.46        & 1.05 $\pm$ 0.12 & 0.6 $\pm$ 0.09 & 1.10 & 0.74 \\ 
$L_2^r$ & 0.71 $\pm$ 0.27 & 1.49        & 1.32 $\pm$ 0.03 & 1.22 $\pm$ 0.08& 1.11 & 1.04 \\
$L_3$   & -2.72$\pm$1.12  & -3.18       & -4.53 $\pm$ 0.14&-3.02$\pm$0.06  &-4.03 &-3.12 \\ 
$L_4^r$ & 0   (fixed)     & 0 (fixed)   & 0.53 $\pm$ 0.39 & 0 (fixed)      &-0.06 & 0.00  \\
$L_5^r$ & 0.91 $\pm$ 0.15 & 1.46        & 3.19 $\pm$ 2.40 & 1.9 $\pm$ 0.03 & 1.34 & 1.26 \\
$L_6^r$ & 0 (fixed)       &   0 (fixed) &      -          & -0.07$\pm$0.20 & 0.15 &-0.01 \\ 
$L_7$   & -0.32$\pm$ 0.15 &-0.49        &       -         & -0.25$\pm$0.18 &-0.43 &-0.49 \\ 
$L_8^r$ &0.62$\pm$ 0.20   & 1.00        &       -         & 0.84 $\pm$ 0.23& 0.94 & 1.06 \\ 
\hline
$L_8^r+2L_6^r$ & 0.62$\pm$ 0.20   & 1.00  & 3.66$\pm$1.52 & 0.7 $\pm$ 0.46 & 1.24 & 1.04 \\
$2L_1^r-L_2^r$ & 0.35$\pm$0.57 &  -0.57 &0.78$\pm$0.24 &-0.02$\pm$ 0.20 & 1.09 & 0.44 \\
\hline
\hline
\end{longtable}
\end{widetext}

In this work, we are interested in the quark mass dependence of the amplitudes,
which appears in ChPT through Lagrangian terms that contain the quark mass matrix ${\cal M}={\rm diag}(\hat m,\hat m, m_s)$, 
that is treated as a perturbation. Note that we work
in the isospin limit $\hat m\sim m_u=m_d=(m_u^{\rm phys}+m_d^{\rm phys})/2$.
Chiral symmetry is explicitly broken by these mass terms and the NGB
acquire masses, which, at leading order, read \cite{Gasser:1984gg}:
\begin{eqnarray}
 M_{0\,\pi}^2&=& 2 \hat m B_0, 
\nonumber \\
M_{0\,K}^2&=&(\hat m + m_s) B_0, \nonumber \\
 M^2_{0\,\eta} &=& {2\over3}(\hat m+2m_s)B_0.
\label{eq:LOmasses}
\end{eqnarray}
Let us recall that the constant $B_0$
is defined from the values {\it in the chiral limit}
of the chiral condensate and the pion decay constant
as follows: $B_0=-< 0\vert \bar qq\vert 0>_0/f_0^2$.
and thus it carries no quark mass dependence. 
To one loop, there are some corrections, and the physical 
meson masses now read:
 \begin{eqnarray}
M_\pi^2&=& M_{0\,\pi}^2\left[1+\mu_\pi-\frac{\mu_\eta}{3}+\frac{16
M_{0\,K}^2}{f_0^2}\left(2L_6^r-L_4^r\right)\right.\nonumber\\
&+&\left.
\frac{8
M_{0\,\pi}^2}{f_0^2}\left(2L_6^r+2L_8^r-L_4^r-L_5^r\right)
\right], \label{pimass}
\end{eqnarray}
 \begin{eqnarray}
M^2_K&=& M^2_{0\,K}\left[1+\frac{2\mu_\eta}{3}+\frac{8
M_{0\,\pi}^2}{f_0^2}\left(2L_6^r-L_4^r\right)\right.\nonumber\\
&+&\left.\frac{8
M_{0\,K}^2}{f_0^2}\left(4L_6^r+2L_8^r-2L_4^r-L_5^r\right)\right],
\label{kmass}
\end{eqnarray}
 \begin{eqnarray}
M^2_\eta&=& M^2_{0\,\eta} \left[1+2\mu_K-\frac{4}{3}\mu_\eta+
\frac{8M^2_{0\,\eta}}{f_0^2}(2L_8^r-L_5^r)\right.\nonumber\\
&+&\left.
\frac{8}{f_0^2}(2 M^2_{0\,K}+M^2_{0\,\pi})(2L_6^r-L_4^r)
\right]\nonumber\\
&+& M^2_{0\,\pi}\left[-\mu_\pi+\frac{2}{3}\mu_K+\frac{1}{3}\mu_\eta\right]
\nonumber\\
&+&\frac{128}{9f_0^2}(M^2_{0\,K}-M^2_{0\,\pi})^2(3L_7+L_8^r),
 \label{etamass}
\\
\mu_P&=&\frac{M_{0\, P}^2}{32 \pi^2 f_0^2}\log\frac{M_{0 \,P}^2}{\mu^2},
\qquad P=\pi,K,\eta. \nonumber
\end{eqnarray}

Note,  however, that all the 
quark mass dependence always appears through the leading order
masses $M_{0\, P}^2$ defined in Eq.\eqref{eq:LOmasses}.
As a matter of fact, this also happens in the ChPT amplitudes,
which means that 
studying the quark mass dependence, keeping $B_0$ fixed,
is nothing but
studying the meson mass dependence.
In practice, and in order to get rid of the $B_0$ constant, we will
recast all our results in terms of masses normalized to
their physical values:
\begin{eqnarray}
&&\frac{\hat m}{\hat m_{\rm phys}}=\frac{M_{0\,\pi}^2}{M_{0\,\pi\,\rm phys}^2}, \label{eq:mhatratio0}\\
&&\frac{ m_s}{m_{s\,\rm phys}}=\frac{M_{0\,K}^2-M_{0\,\pi}^2/2}{M_{0\,K\,\rm phys}^2-M_{0\,\pi\,\rm phys}^2/2}.
\label{eq:msratio0}
\end{eqnarray}
Note that, from now on, a quantity with a ``phys'' subscript refers to the 
value of that quantity in the physical case.
Thus, in this work we will change quark masses, 
that, using Eqs.\eqref{pimass},
\eqref{kmass} and \eqref{etamass}, imply a change in meson masses, which are the ones appearing explicitly in the ChPT scattering amplitudes.  

\begin{figure}
  \centering 
  \includegraphics[scale=.7]{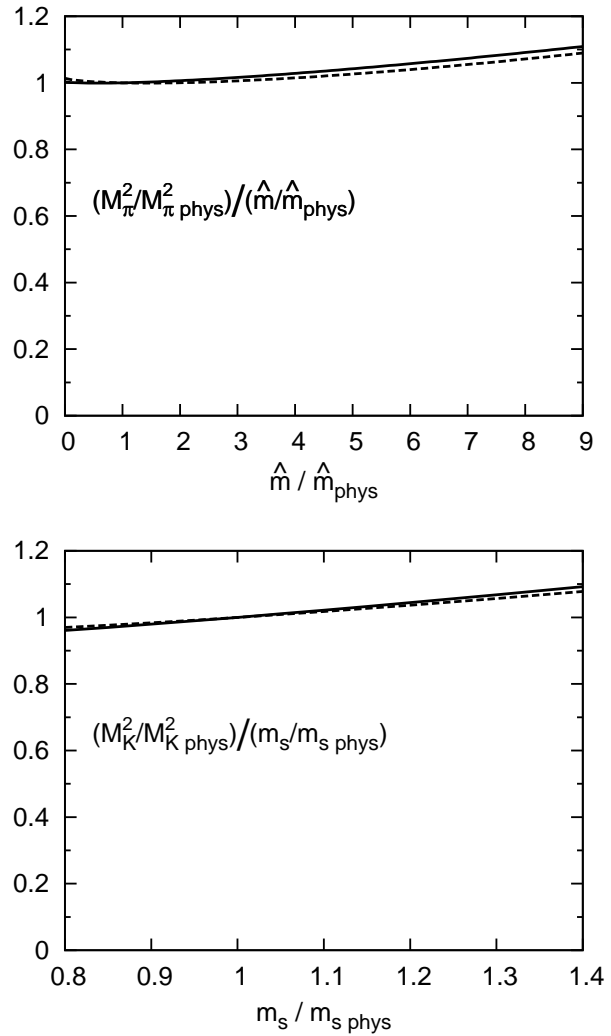}
  \caption{ \label{fig:masses}
Top: The ratio $(M_\pi^2/M_{\pi\,\rm phys}^2)/(\hat m/ \hat m_{\rm phys})$
Bottom:
$(M_K^2/M_{K \,\rm phys}^2)/(m_s/m_{s\,\rm phys)}$.
Within the range of variation of this work, 
a relative variation of a quark mass can be also understood
as  the same relative variation in 
the corresponding meson mass squared to within 
$\sim 10\%$ accuracy.
The continuous and dashed lines correspond to Fit I and II sets of LECs given in
Table~\ref{tab:LECs}.
}
\end{figure}

There are many advantages in using meson masses
as the variation parameter, since, contrary to quark masses
that have a complicated and scale dependent definition 
on the QCD renormalization scheme, meson masses are
observables, with no scale dependence and a straightforward physical 
interpretation. Actually, many lattice results are also recast 
in terms of pion or kaon mass variations. 
Unfortunately the simple relations in Eqs.\eqref{eq:mhatratio0}
and \eqref{eq:msratio0} are exact only when 
written in terms of the leading order masses $M_{0\,P}$, not
the observable ones.
Nevertheless, the one-loop corrections become numerically small 
when taking ratios so that, to a good degree of approximation,
the reader still can think in terms of physical meson masses
instead of their leading order values.
Actually, in Fig.\ref{fig:masses} we show that 
within the range of quark mass variations that we will consider in this work, 
the naive, but intuitive, 
relations
\begin{eqnarray}
&&\frac{\hat m}{\hat m_{\rm phys}}
\simeq \frac{M_{\pi}^2}{M_{\pi\,\rm phys}^2}, \quad {\rm keeping}\; m_s= m_{s\,\rm phys}, \\
&&\frac{ m_s}{m_{s\,\rm phys}}
\simeq\frac{M_{K}^2}{M_{K\,\rm phys}^2}, \quad {\rm keeping}\;\hat m= \hat m_{\rm phys} 
\end{eqnarray}
are a very good approximation --
within less than 10\% error-- to the correct ratios 
in Eqs.(\ref{eq:mhatratio0}) and (\ref{eq:msratio0}),
that we actually use. To make our presentation of the results
more intuitive we will give, when possible,
our results both in terms of quark mass variation 
and the corresponding meson mass variation.

At this point we have to address the question of how much we can vary the
quark masses before our approach breaks down. 
First we want the pion 
always lighter than the kaon and eta since otherwise the elastic 
approximation
would make no sense for $\pi\pi$ or $K\pi$ scattering.
Second, ChPT  seems to work for masses 
as high as 500 MeV, since we already know
that it provides a fairly good
description of low energy $K \pi$ scattering, 
even though $M_K\sim 500 \,\rm MeV$.
Thus, when changing the non strange quark mass, keeping $m_s$ fixed,
we will show results up to 
$M_\pi< 440\,\rm MeV$ but not beyond, since then 
$M_K\simeq 600\,\rm MeV$.
Equivalently,  this means $\hat m/\hat m_{\rm phys}\leq 9$.
Concerning the strange quark variation with $\hat m$ fixed, 
we will consider 
$0.7<m_s/m_{s\,\rm phys}<1.3$, since $M_\pi$ barely changes and 
$400\,\rm MeV<M_K<585\,\rm MeV$. This ensures that the
$m_K+m_\eta$ is not below the $m_{K^*}$ mass
so that we would need a coupled channel formalism.
Of course, the closer to the estimated applicability limits
 the less reliable 
our formalism will be.

The SU(3) $\pi\pi$ and $K\pi$ 
one-loop amplitudes were first calculated in \cite{Kpi},
although for technical reasons explained in \cite{GomezNicola:2001as}
needed for the implementation of exact unitarity later on, 
we use the expressions in the appendix of \cite{GomezNicola:2001as}, 
but written in terms
of all physical constants $M_\pi, M_K, M_\eta$, $f_\pi, f_K, f_\eta$
as explained in \cite{Pelaez:2004xp}. For completeness
we show here the decay constant dependence on meson masses.
\begin{eqnarray}
f_\pi&=& f_0\left[1-2\mu_\pi-\mu_K+\frac{4
M_{0\,\pi}^2}{f_0^2}\left(L_4^r+L_5^r\right)+\frac{8
M_{0\,K}^2}{f_0^2}L_4^r\right], \nonumber\\ f_K&=&
f_0\left[1-\frac{3\mu_\pi}{4}-\frac{3\mu_K}{2}-\frac{3\mu_\eta}{4}+\frac{4
M_{0\,\pi}^2}{f_0^2}L_4^r\right.\nonumber\\&+&\left.\frac{4
M_{0\,K}^2}{f_0^2}\left(2L_4^r+L_5^r\right)\right],
 \label{fpis}\\
f_\eta&=&f_0\left[1-3\mu_K+ \frac{4
L_4^r}{f_0^2}\left(M_{0\,\pi}^2+2M_{0\,K}^2\right)
+\frac{4M_{0\,\eta}^2}{f_0^2}L_5^r\right].
\nonumber
\end{eqnarray}

Of course, for $\pi\pi$ and $K \pi$ elastic scattering the most relevant
quark mass dependence comes via $M_\pi$, $M_K$ and $f_\pi$, $f_K$ 
(since etas only appear in loops). Consequently, the LECs that play the most important
role are $L_4, L_5, L_6$ and $L_8$, since they  appear in Lagrangian terms that 
contain explicitly powers of the quark mass matrix.
In contrast, the Lagrangian terms proportional to the $L_1, L_2$ and $L_3$ 
constants only contain derivatives and thus are somewhat less relevant for
the quark mass dependence, but more relevant in terms of $s$ dependence.

Finally, let us remark that despite the fact that their effect is encoded in the LECs,
the ChPT amplitudes, being an expansion, cannot describe
resonances and their associated poles in the second Riemann sheet.
Actually, resonances are usually identified with a saturation 
of the unitarity constraints, which for elastic partial waves
of definite isospin $I$ and angular momentum $J$
read: 
\begin{equation}
\ima t_{IJ}(s) = \sigma(s) \vert 
t_{IJ}(s)\vert^2 \; \Rightarrow \;  
\vert t_{IJ}(s)\vert \leq 1/\sigma(s),
\label{unit}
\end{equation}
where $\sigma(s)=2k/\sqrt{s}$ and 
$k$ is the center of mass momentum.
The above equations imply that the partial wave
can be recast in terms of a single phase or ``phase shift'':
\begin{equation}
t_{IJ}(s)=\exp{i\delta_{IJ}(s)}\sin\delta_{IJ}(s)/\sigma(s)
\label{phaseshift}
\end{equation}
In this work we are only interested in the $(I,J)=(0,0), (1,1)$
and $(2,0)$ channels for $\pi\pi$ scattering and $(I,J)=(1/2,0), (1/2,1)$
and $(3/2,0)$ for $\pi K$ scattering. For simplicity we will drop
the $IJ$ subindex when discussing general properties of 
elastic partial waves.

Note, however, that the ChPT  
expansion Eq.\eqref{eq:ChpTseries}, being basically
a polynomial in energy, violates
the bound in Eq.\eqref{unit} as the energy increases and cannot generate poles.
Still, ChPT satisfies elastic unitarity 
perturbatively:
\begin{equation}
\ima t_2(s)=0, \quad \ima t_4(s)=\sigma\vert t_2(s)\vert^2, ... \label{eq:pertunit}
\end{equation}
But, of course, elastic unitarity can be badly
violated if the ChPT series is extrapolated close to a resonance.
For these reasons, the resonance region lies beyond
the reach of standard ChPT. However, we will see next that
ChPT can be used in an alternative way.

\subsection{Dispersion relations, unitarity and ChPT}

Instead of 
simply extrapolating
its series to higher energies, ChPT can be used to calculate the subtraction 
constants of a dispersion
relation for the two-body amplitude. 
These constants correspond to the values of the amplitude
or its derivatives at a low energy point where 
the use of ChPT is well justified. The remaining information
to build the amplitude comes from the strong constraints
of analyticity and unitarity.

First of all, it is straightforward to 
rewrite the strong non-linear elastic unitarity constraint
given in Eq.\eqref{unit}, as follows
\begin{equation}
  \ima 1/t(s)=-\sigma(s).
\label{eq:unitinverse}
\end{equation}
This means that, from unitarity, 
 we know {\it exactly} the imaginary part of $1/t$
in the elastic region. We are only left to determine the real part of $1/t$.

Concerning the analyticity constraints, for simplicity
let us consider first the case of two 
identical particles, as in $\pi\pi$ scattering.
Then, the analytic structure in the complex
$s$ plane is rather simple: it has a ``right'' or ``physical'' 
cut  on the real axis from 
threshold to $+\infty$, and a ``left cut'' from 
$-\infty$ to $s=0$. By means of the Cauchy Theorem, a dispersion relation 
provides the amplitude anywhere 
inside the cut complex plane
in terms of a weighted integral of its imaginary part over the
cuts. 

In our case, instead of $t$ we are interested
in a dispersion relation for $1/t$ since we know
exactly its imaginary part in the elastic region thanks to
Eq.\eqref{eq:unitinverse}. For convenience, and since
$t_2$ is real, instead of $1/t$ we define 
$G=t_2^2/t$, that also has a right cut ($RC$) and
a left cut ($LC$). Since scalar waves are known to 
have dynamical Adler zeros in the low energy region below threshold,
we will also allow for a pole contribution $PC$ in $G(s)$.
All in all, we can write a dispersion relation for $G(s)$ as follows
\begin{eqnarray}
  \label{disp1/t}
  G(s)&=&G(0)+G'(0)s+\tfrac{1}{2}G''(0)s^2\\&+&
  \frac{s^3}{\pi}\int_{RC}ds'\frac{\ima G(s')}{s'^3(s'-s)}+
  LC(G)+PC.\nonumber
\end{eqnarray}
In the elastic region, unitarity in Eq.\eqref{unit}, together with
Eq.\eqref{eq:pertunit}, 
allow us to evaluate \emph{exactly} 
$\ima G=-\sigma t_2^2=-\ima t_4$ on the $RC$. Note the three $1/s'$ factors --
called subtractions -- that we have introduced to suppress the 
high energy part and in particular the inelastic 
contributions, so that the integrals are dominated by the 
low energy region. But once the integrals are dominated by the low energy,
it is well justified to use ChPT inside the integrals and thus, for instance,
the $LC$ integral to one loop ChPT is given by
$LC(G)\simeq LC(-t_4)+...$. 

The price to pay for the three subtractions is that
analyticity only determines the function up to a second order polynomial 
$G(0)+G'(0)s+\tfrac{1}{2}G''(0)s^2$. However, note that its coefficients
correspond to the values of the amplitude or its derivatives at $s=0$,
where ChPT can be safely applied. In particular, to one-loop,
$G(0)\simeq t_2(0)-t_4(0)$, $G'(0)\simeq t_2'(0)-t_4'(0)$ and $G''(0)=-t_4''(0)$, since $t_2''(0)$ vanishes.
Let us neglect for the moment the pole contribution, 
which is of higher order and
only numerically relevant below threshold. 
Then one finds that all contributions can be recast in terms of the leading
$t_2(s)$ and next to leading $t_4(s)$ ChPT amplitudes. 
Finally, we arrive at 
the so-called Inverse Amplitude Method (IAM)
 \cite{Truong:1988zp,Dobado:1996ps}:
\begin{equation}
  \label{eq:IAM}
  t(s)\simeq \frac{t_2^2(s)}{t_2(s)-t_4(s)}\,.
\end{equation}
Remarkably, this simple equation ensures elastic unitarity, matches ChPT
at low energies, and,  using LECs 
compatible with existing determinations,
describes fairly well data up to somewhat 
less than 1 GeV, generating
the $\rho$, $K^ *$, $\sigma$ and $\kappa$ resonances
as poles on the
second Riemann sheet.
It has been shown \cite{Oller:1997ti} 
that the scalars can actually be generated mimicking
the LEC, tadpole and crossed channel diagrams by a cutoff of natural size,
and thus it is said that scalars are ``dynamically generated'' from,
essentially, meson-meson dynamics (meson loops). In contrast,
to generate the vectors, a precise knowledge of the LECs is needed, namely,
of the underlying, non meson-meson QCD dynamics.

Here we will update this description of experimental
data but furthermore
we will simultaneously  describe the existing 
lattice results for decay constants and some scattering lengths.

The IAM equation above is just the one-loop result,
but it can be easily and systematically
extended to higher orders of ChPT or
generalized within a coupled channel formalism 
\cite{Oller:1997ng,GomezNicola:2001as,Pelaez:2004xp}, 
generating also the $a_0(980)$, $f_0(980)$ and the octet $\phi$.
However note that there is no dispersive justification for the
coupled channel approach formula \footnote{ If we followed a similar approach
the left cuts would mix when calculating 
the inverse matrix
and produce spurious analytic structures}
and that is the main reason, apart from simplicity,
why we have restricted our analysis to the elastic case.

For completeness, and even though it will be 
negligible except for very high masses near the
applicability limits of our approach, 
let us now include the pole contribution $PC$
ignored so far. Its contribution can be calculated explicitly from
its residue  \cite{GomezNicola:2007qj} and, to one-loop,  we
find a modified IAM (mIAM) formula:
\begin{eqnarray}
  \label{mIAM}
  t^{mIAM}&=& \frac{t^2_2}{ t_2-t_4 +A^{mIAM}}\\
  A^{mIAM}&=&t_4(s_2){-}\frac{(s_2{-}s_A)(s{-}s_2)
    \left[t'_2(s_2){-}t'_4(s_2)\right]}{s{-}s_A},\nonumber
\end{eqnarray}
where $s_A$ is the position of the Adler zero in the $s$-plane, and
$s_2$ its LO approximation.
The standard IAM
is recovered for $A^{mIAM}=0$, which holds exactly for
all partial waves except the scalar ones. Above, and in the usual IAM
derivation \cite{Dobado:1996ps} $A^{mIAM}$ was neglected,
since it formally yields a NNLO contribution and  is numerically
very small, except near the Adler zero, where it diverges. However,
if $A^{mIAM}$ is neglected, the IAM Adler zero occurs at $s_2$,
correct only to LO, it is a double zero instead of a simple one, and
a spurious pole of the amplitude
appears close to the Adler zero. All of these caveats are removed
with the mIAM, Eq.~\eqref{mIAM}. The differences in the 
physical and resonance region between the IAM and the mIAM are less
than 1\%. However, as we will see, for large $M_\pi$ the $\sigma$ and $\kappa$
poles ``split'' into two virtual poles below threshold, 
one of them moving towards
zero and approaching the Adler zero region, where the IAM fails. Thus, we will
use for our calculations the mIAM, although it is only relevant for the 
mentioned second $\sigma$ and $\kappa$ poles, and only when 
they are very close to their corresponding Adler zeros.

Finally, we want to comment on the unequal mass case, since we also 
want to describe $K\pi$ elastic scattering. The main difference now is that
the left cut extends from 
$-\infty$ to $s=(M_1-M_2)^2$, and also that there is a circular cut, 
centered at $s=0$ with radius $\vert M_1^2-M_2^2 \vert$. Again their main 
contribution comes from a region where ChPT can be applied. This time, however,
$t_2(s)$ has
\emph{two} zeros instead of one, $s_{2\pm}=\frac1{5}\left(M_K^2+M_\pi^2\pm
2\sqrt{4M_K^4-7M_K^2M_\pi^2+4M_\pi^4}\right)$, and the modification
to the IAM reads:
\begin{widetext}
\begin{equation}
  A^{mIAM}(s)= \frac{t_2(s)^2}{t_2'(s_{2+})^2}
  \left[
    \frac{t_4(s_{2+})}{(s-s_{2+})^2}-
    \frac{(s_{2+}-s_A)}{(s-s_{2+})(s-s_A)}
    \left(
      t'_2(s_{2+})-t_4'(s_{2+})+
      \frac{t_4(s_{2+})t''_2(s_{2+})}{t'_2(s_{2+})}
    \right)
  \right].
\label{Aangelpik}
\end{equation}
\end{widetext}
Once again we note that this modification will be numerically negligible except
in the close vicinity of the Adler zero. The poles of the resonances 
under study will only come close to that region for very high values 
of the quark masses,
in the limit of applicability of ChPT and our approach.

Before describing our fits, 
 we want to remark that, in the IAM derivation above,
ChPT does not play any role outside its applicability limits.
By including three subtractions we have suppressed strongly all 
contributions to the integrals in high energy regions where ChPT results 
are not reliable. Finally, the three subtraction constants, which correspond 
to values of the amplitudes or their derivatives at $s=0$ are  well
calculated with ChPT. Of course, this is just a one-loop calculation, 
although the generalization to higher orders is tedious but straightforward.
Hence, our approach does not model the left or inelastic cuts,
but just uses the corresponding ChPT approximation
that, in principle, can be improved order by order -- eventually
 including more subtractions.

\section{Fits to data and lattice results}
\label{sec:fits}

As commented before, it has been known for long \cite{Dobado:1996ps} that with
the one-loop elastic IAM (the mIAM is almost identical) in Eq.\eqref{eq:IAM}
it is possible to obtain a remarkable description of
$\pi\pi$ and $K\pi$ experimental data up to somewhere below 1 GeV. 
Simultaneously, the IAM generates the
poles associated to the $f_0(600)$, $\rho(770)$, $K^*(892)$
and $\kappa(800)$ resonances and this is achieved using parameters 
compatible with those of standard ChPT \cite{GomezNicola:2001as}.
However, that description was obtained from a fit to experimental data,
and therefore it is mostly sensitive to the LECs  $L_1,L_2,L_3$
that predominantly  govern the $s$ dependence of partial waves, 
but much less so to the rest of LECs that carry an explicit meson mass
dependence.
Of course, since now 
 we want to extrapolate the IAM fits to non-physical masses,
it is very important that we use a good description of the  mass dependence
in observables like masses, decay constants, etc. 
before extracting conclusions about resonance behavior.
For that reason, we are presenting here an updated IAM description of
experimental data simultaneously fitted to the available lattice results on 
the mass dependence of $M_\pi/f_\pi$, $M_\pi/f_K$ and $M_K/f_K$ as well as 
scattering lengths for the doubly charged channel in $\pi\pi$, $K\pi$ and
$KK$ scattering. Note that for the moment, these lattice data are only available in the highest isospin combination for each particle pair.

\begin{figure*}
  \centering
  \includegraphics[scale=.57,angle=-90]{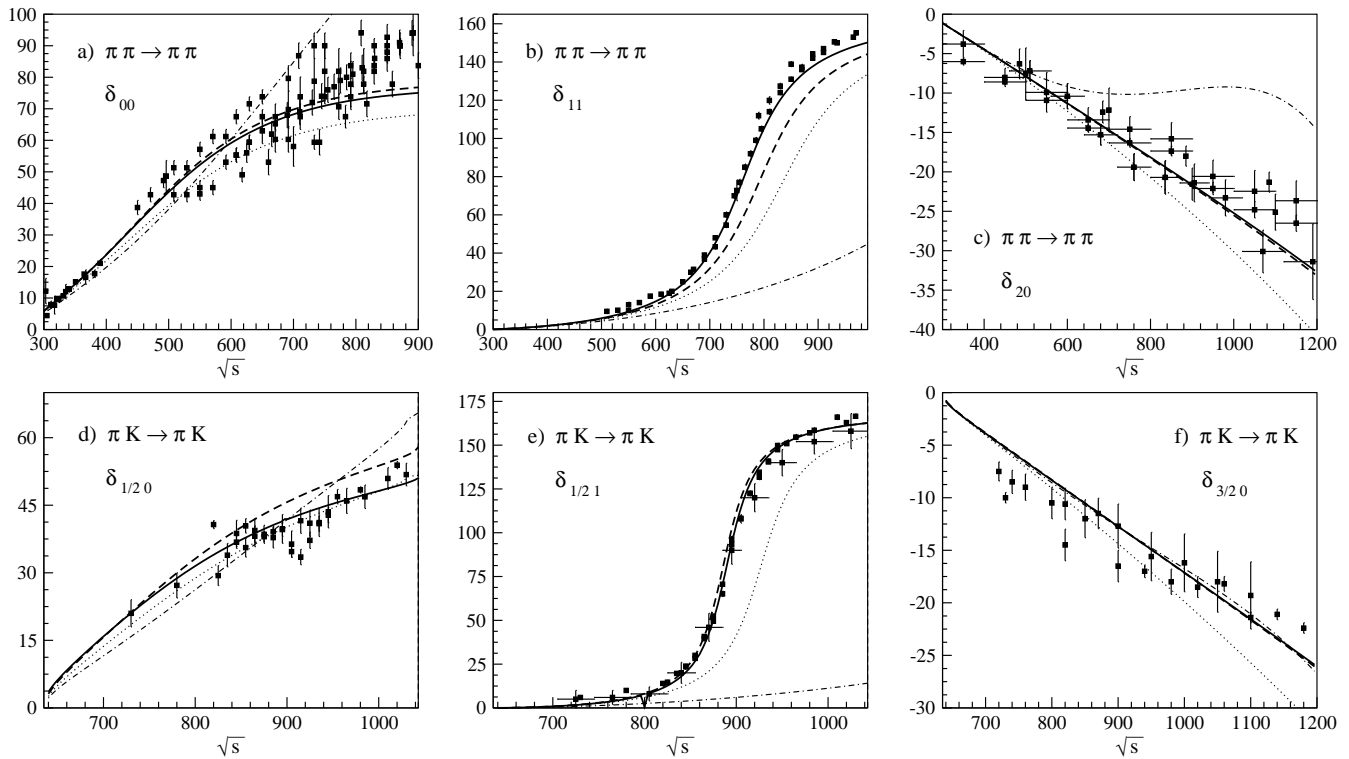}
\parbox{\linewidth} { 
\caption{ \label{fig:experiment}
Results of our IAM fits versus experimental data on $\pi\pi$ and $\pi K$
scattering. The continuous and dashed lines correspond, respectively, to Fits I and II,
whose parameters are given in Table \ref{tab:LECs}. For comparison
we show the results of the IAM 
if we used the ChPT LECs obtained from the two-loop analysis of $K_{l4}$
decays listed also in Table \ref{tab:LECs} (dotted line) as well as the results of
standard non-unitarized ChPT with the same set of LECs (dot-dashed line). 
The plotted data correspond to experimental results \cite{experimentaldata}, 
which are often incompatible. For that reason, in our fits, 
and for $\pi\pi$ scattering, we have actually
used the results of a dispersive analysis of these data \cite{Kaminski:2006qe}.
} 
}
\end{figure*}

In order to change the masses and 
decay constants according to Eqs.\eqref{pimass},\eqref{kmass},\eqref{etamass}
and \eqref{fpis}, 
we need first to extract the tree level quantities:
$M_{0\pi}^{2}$, $M_{0K}^2$ and $f_0$ from
the physical values of the pion, kaon and eta  masses 
as well as the three decay constants $f_\pi, f_K$ and $f_\eta$. 
Note that  $M_{0\eta}^2$ will be obtained
from the Gell-Mann-Okubo relation: 
$4M_{0K}^2-M_{0\pi}^{2}-3M_{0\eta}^2=0$. 
Since there are more physical values than 
tree level constants, for a given set of 
LECs we actually use the tree level 
constants that best fit the physical ones. Thus, 
the physical masses and decay 
constants that we will obtain when recovering 
them from the tree level ones will 
be only approximate. This is, of course, the consequence of using
a truncated expansion -- ChPT to one-loop -- to describe observables.

We have made two fits whose resulting LECs sets are given in
the two last columns of 
Table~\ref{tab:LECs}. 
Since there are many parameters, there are strong correlations. 
Thus, sets with quite different parameters can give raise to acceptable
descriptions of data, depending on how one weights experiment and lattice
results. On Fit I we have fitted to experimental data 
coming from \cite{experimentaldata} and
to lattice results given in \cite{lattice}. 
We show in Fig.1 these data, for 
the $(I,J)=(0,0)$ and $(2,0)$ waves,
 where many different experiments are actually incompatible, but we have fitted to 
 the phase shifts arising from the dispersive
analysis of the experimental
data  in \cite{Kaminski:2006qe}, where a complete 
set of Forward Dispersion Relations and Roy Eqs. was constrained
on a phenomenological fit to all waves. For the $(1,1)$ wave
we have used also the phenomenological phase shifts from that solution since,
apart from the dispersive constraints,
it fits the data of the electromagnetic form factor of the pion,
which is much more reliable and precise than the existing experiments
on $(1,1)$ pion-pion scattering.
Anyway, since for $\pi K$ and other waves we are still using scattering,
and also because the method has an intrinsic error due to the NLO approximation
on the integrals, we have added
in quadrature to the experimental
 data errors a constant error of 2 degrees and a variable error of 5\% 
of the phase shift and to lattice 
results on masses over decay constants 5\% of their values also 
in quadrature to their errors. We have 
also introduced a constraint so that the LECs don't differ
much from those found in the $K_{l4}$ analysis 
to two loops of \cite{Amoros:2001cp}, by weighting also in the $\chi^2$
the LECs with the values in \cite{Amoros:2001cp}. 
On Fit II we have given an additional weight to 
the large $1/N_c$ constraint $2L_1-L_2=0$ (dividing its error by 10 when
calculating the $\chi^2$) whereas we have 
relaxed the constrains on $\delta_{11}$ and $\delta_{1/2\, 0}$ 
(dividing their $\chi^2$ by $1.5$).  

For comparison, also in Table~\ref{tab:LECs}
 we provide three typical sets of LECs available in the literature
obtained from 
data analyses using dispersive techniques plus ChPT.
Those on the first and the second columns, come from a one and two-loop
analysis of $K_{l4}$ decays
\cite{Amoros:2001cp}, where $L_4$ and $L_6$
were set equal to zero (following leading order $1/N_c$ arguments).
The ``Roy-Steiner'' column comes from 
a dispersive analysis of $\pi K$ scattering \cite{Buettiker:2003pp}. 
Note that the LECs in these sets are frequently within 
more than two standard deviations from one another, and we consider
that their difference is indicative of the typical size
of systematic uncertainties in our knowledge of LECs.
As commented above, since the one-loop IAM generates
correctly only the s-channel leading logs of the two-loop calculation, 
which are dominant at low energies, it is not clear whether
we should compare with the LECs obtained in the one or two loop
ChPT analysis. Actually,  all of our 
IAM LECs lie very close, or within the uncertainties,
of at least one of the previous determinations given in the Table. 
Taking into account the uncertainties in these non-unitarized
determinations, we consider that the agreement between the IAM LECs
 and previous determinations is fair.
Let us remark that the relevant fact 
about this comparison is to note that we do not 
need to make any fine tuning of the LECs,
 like changing well established signs, changing order of magnitude, etc. ,
to describe the experimental and lattice data simultaneously.

Finally, we also provide in Table~\ref{tab:LECs}, the IAMIII set of LECs,
which corresponds to one of the three fits 
obtained using the coupled channel IAM in  \cite{Pelaez:2004xp}.
This set was fitted to experimental data only and the uncertainties
quoted are just statistical. Taking into account that we 
are using the single channel IAM instead of the coupled channel one,
and the estimate
of systematic uncertainties discussed above, we see that our new
fits including new experimental data and lattice results,
 are not too different from those already obtained in \cite{Pelaez:2004xp}.

In Fig.\ref{fig:experiment} we show the results of our fits
 compared with experimental 
data on $\pi\pi$ and $\pi K$ elastic scattering phase shifts. 
The best description is given by Fit I (continuous line), 
whereas Fit II gives a somewhat too heavy
$\rho(770)$ vector resonance (by roughly 50 MeV, i.e., a 6\% error). For 
comparison, we show as a dotted line the results of the IAM 
if we used the ChPT LECs obtained from the two-loop analysis of $K_{l4}$
decays listed in Table \ref{tab:LECs}.
We also show as a dot-dashed line the results that would be obtained if 
the non-unitarized ChPT one loop results are extrapolated to higher energies 
using the same set of LECs. Note that the IAM
results describe rather well both the resonant and non resonant shapes up to 1 GeV 
or slightly above, except for the scalar isoscalar $\delta_{00}$,
that is only described up to 800 MeV. This is due to the presence of the sharp
rise caused by the $f_0(980)$ resonance, that decays mostly to two kaons and
can only be described with the coupled channel IAM formalism, 
\cite{Oller:1997ng,GomezNicola:2001as,Pelaez:2004xp}, 
that we do not use here
for the reasons explained above.

\begin{figure*}
   \centering
   \includegraphics[scale=.58,angle=-90]{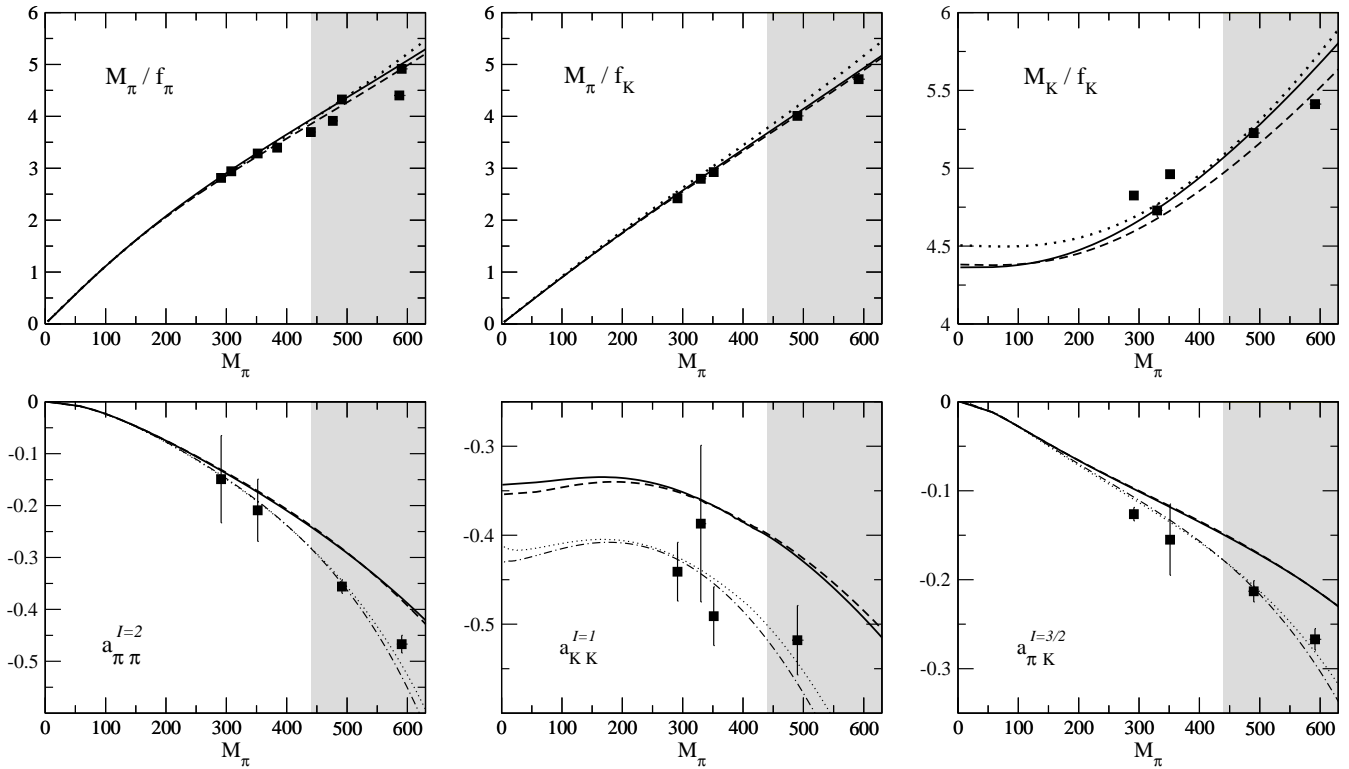}
\parbox{\linewidth} { 
\caption{ \label{fig:latticedata} 
Result of the unitarized fits to lattice calculations of $M_\pi/f_\pi$, $M_\pi/f_K$,
$M_K/f_K$ and the $\pi^+\pi^+$, $K^+ K^+$, $K^+\pi^+$ 
scattering lengths. The continuous and dashed lines correspond, respectively, to Fits I and II,
whose parameters are given in Table \ref{tab:LECs}. For comparison
we show the results of the IAM if we used the ChPT LECs obtained from the two-loop analysis
of $K_{l4}$ decays listed also in Table \ref{tab:LECs} (dotted line) as well as the results of
standard non-unitarized ChPT with the same set of LECs (dot-dashed line). 
Lattice results come from \cite{lattice}. The grey area lies beyond 
our applicability region, however, it is useful to check that
 our description does not deteriorate too rapidly.
}
}
\end{figure*}

Those results are, of course, well known, 
and these fits would just be an update of \cite{Pelaez:2004xp} if
we had not also included lattice data on 
the fit, that we show in Fig.\ref{fig:latticedata}.
Note that we are fitting results on  $M_\pi/f_\pi$, $M_\pi/f_K$ and $M_K/f_K$ 
and the $\pi^+\pi^+$, $K^+ K^+$, $K^+\pi^+$ scattering lengths \cite{lattice}. 
Once again we show Fits I and II as continuous and dashed lines, respectively,
together with IAM results using the LECs from the two-loop analysis of $K_{l4}$
decays listed in Table \ref{tab:LECs} (dotted line) and 
non-unitarized ChPT to one-loop with the same set of LECs (dot-dashed line).
As explained above, we do not consider that our method should be trusted
for pion masses heavier than 440 MeV, being optimistic, and that is 
why the heavier mass region is shown as a grey area.

\section{Dependence on $\text{\large \it u}$ and $\text{\large \it d}$ quark masses}

Now that we have a good description of both the 
energy dependence of pion-pion amplitudes together with
the mass dependence of the few observables available from lattice,
we can change the value of the light quark mass, keeping $m_s$ fixed,
 and predict the 
behavior of the resonances generated within the IAM.

\subsection{Light vector mesons: the $\rho(770)$ and $K^*(892)$}

The $\rho(770)$ and $K^*(892)$ vector
resonances are well established $q \bar q$ states belonging to 
an SU(3) octet. The first is produced in $\pi\pi$ scattering,
and its quark mass dependence was already studied within SU(2) ChPT
\cite{Hanhart:2008mx}. Here we will just check
 that we reobtain very similar results within the SU(3) formalism,
while describing simultaneously the lattice observables shown in Fig.\ref{fig:latticedata}. However, the $K^*(892)$ appears in $\pi K$ scattering and 
can only be obtained using SU(3) ChPT as we do here. 

\begin{figure*}
  \centering
  \includegraphics[scale=.73]{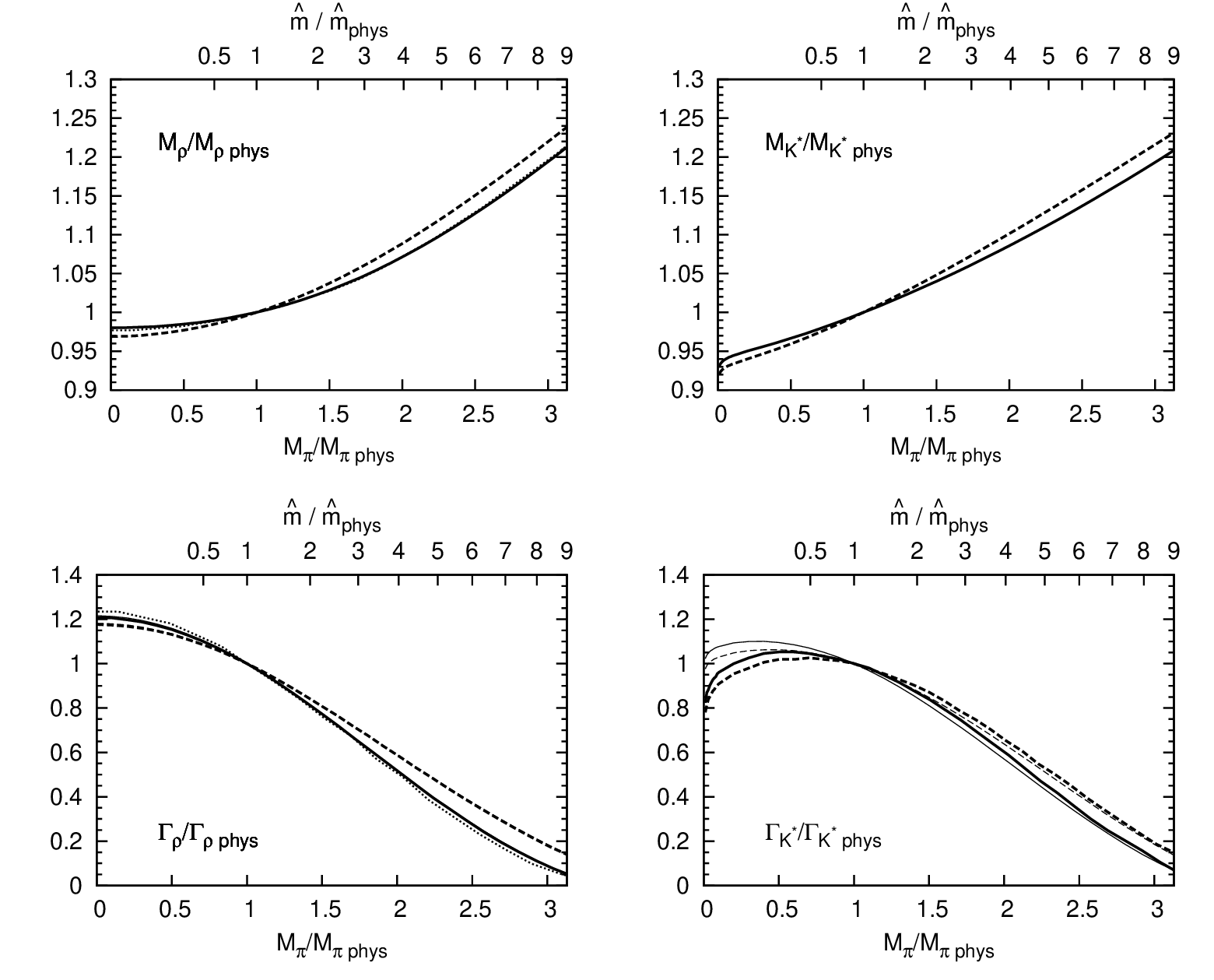}
\parbox{\linewidth} { 
    \caption{ \label{fig:rhoksmhatdepend} 
Dependence of the $\rho(770)$ 
 and $K^*(892)$ mass and width with respect to
the non-strange quark mass $\hat m$  (horizontal upper scale), or
the pion mass (horizontal lower scale).
Note that we give all quantities normalized to their 
physical values. 
The thick continuous and dashed lines correspond, respectively, to
Fit I and Fit II described in the text with unitarized SU(3) ChPT.
For the $\rho$ these results are very 
compatible with those in \cite{Hanhart:2008mx}
using  SU(2) ChPT (dotted line).
The continuous (dashed) thin line shows the $M_\pi$ dependence of
the widths from the change of phase space only, assuming a constant
coupling of the resonances to two mesons, $\rho(770)$ to $\pi\pi$ and 
$K^*(892)$ to $\pi K$, calculated from the
dependence of masses and momenta given by Fit I (II). For the $\rho(770)$
the thin and dashed lines overlap completely.
}
}
\end{figure*}

\subsubsection{Mass and width}

Thus, in Fig.\ref{fig:rhoksmhatdepend}, we show the dependence of the
light vector resonances on the non-strange quark masses, using one-loop
 SU(3) ChPT unitarized with the IAM. For each resonance, 
these masses and widths are defined
from the position of their associated pole in the second Riemann sheet,
through the usual Breit-Wigner identification: $\sqrt{s_{pole}}\equiv M-i\,\Gamma/2$.
We show the results for
Fits I and II as continuous and dashed lines, respectively.
The results for both fits are very consistent and their difference 
can be taken as an estimation for systematic uncertainties in the 
choice of LECs.
To suppress \textit{systematic uncertainties} 
we give all quantities normalized to their physical values.
Note that we provide two scales for the mass variation:
In the upper horizontal axis, we show the variation of the quark
mass in terms of $\hat m/\hat m^{\rm phys}$, whereas
in the lower horizontal axis we show the variation of the pion mass in terms
of $M_\pi/M_{\pi}^{\rm phys}$. The one-loop ChPT relation between these 
two scales is given by Eqs.\eqref{eq:LOmasses} and \eqref{pimass}.
To be precise, this relation changes for different
LECs, but, as we already showed in Fig.\ref{fig:masses}, 
the difference is too small to be observed with the
 naked eye in the axes of Fig.\ref{fig:rhoksmhatdepend}.

In the left panels we also show, as a dotted line,
 the SU(2) ChPT result  already
obtained in \cite{Hanhart:2008mx}, which, is fairly consistent 
with the new SU(3) results. Of course, the difference is somewhat larger 
when the pion mass is closer to the kaon mass, and the kaons start playing
a more prominent role. Of course, since the SU(2) results \cite{Hanhart:2008mx}
already described fairly well the available lattice calculations 
for the $\rho(770)$ mass, so it happens with the SU(3) results here.
In addition, this ensures that the $M_\rho$ dependence on $M_\pi$
 agrees nicely with the estimations 
for the two first coefficients of its chiral expansion \cite{bruns},
which was already checked in the SU(2) case \cite{Hanhart:2008mx}.

Since the vertical scale is the same for the $\rho(770)$ and $K^*(892)$
plots, the similarity of their  behavior 
is very evident. Both their masses increase smoothly 
as the quark mass increases, but much slower than the pion mass.
Some differences can be observed for small $\hat m$, but this is due to the fact that the SU(3) breaking between the $\rho(770)$ and the $K^*(892)$
is more evident since we
keep $m_s$ fixed to its large physical value. 
What is interesting to observe is that the naive rule of thumb frequently
used in the literature \cite{valence}, that 
$\partial M_R/\partial \hat m= N_R^v$, where $N_R^v$ is the number of valence
non-strange quarks, yields the correct order of magnitude,
(and this is how it has been used in \cite{valence})
but would predict a 2:1 relation  for the slope of the $\rho(770)$
with respect to that of the $K^*(892)$, which is not observed 
for light quarks. 

Continuing with our analysis, we note
that, as the quark mass increases,
 the two-pion and pion-kaon threshold grow faster than 
the masses of the resonances and,
as a consequence, there is a strong phase space suppression than can account
exclusively for the decrease of their widths.
We show in the lower panels the $M_\pi$ 
dependence of $\Gamma_\rho$ and $\Gamma_{K^*}$ normalized to their physical values.
The decrease of the widths is largely kinematical, following remarkably well
the expected reduction from phase space as the masses of the NGB increase
(thin continuous and dashed lines corresponding to Fit I and Fit II respectively,
although for the $\rho(770)$ they overlap so well that the thin lines are not seen ). 
This result was already found for the $\rho(770)$ 
within the SU(2) formalism and is nicely confirmed here.
This suggests that there is no dynamical effect through
the vector coupling to two mesons, as we will analyze next.

\begin{figure*}[t]
  \centering
  \includegraphics[scale=.72]{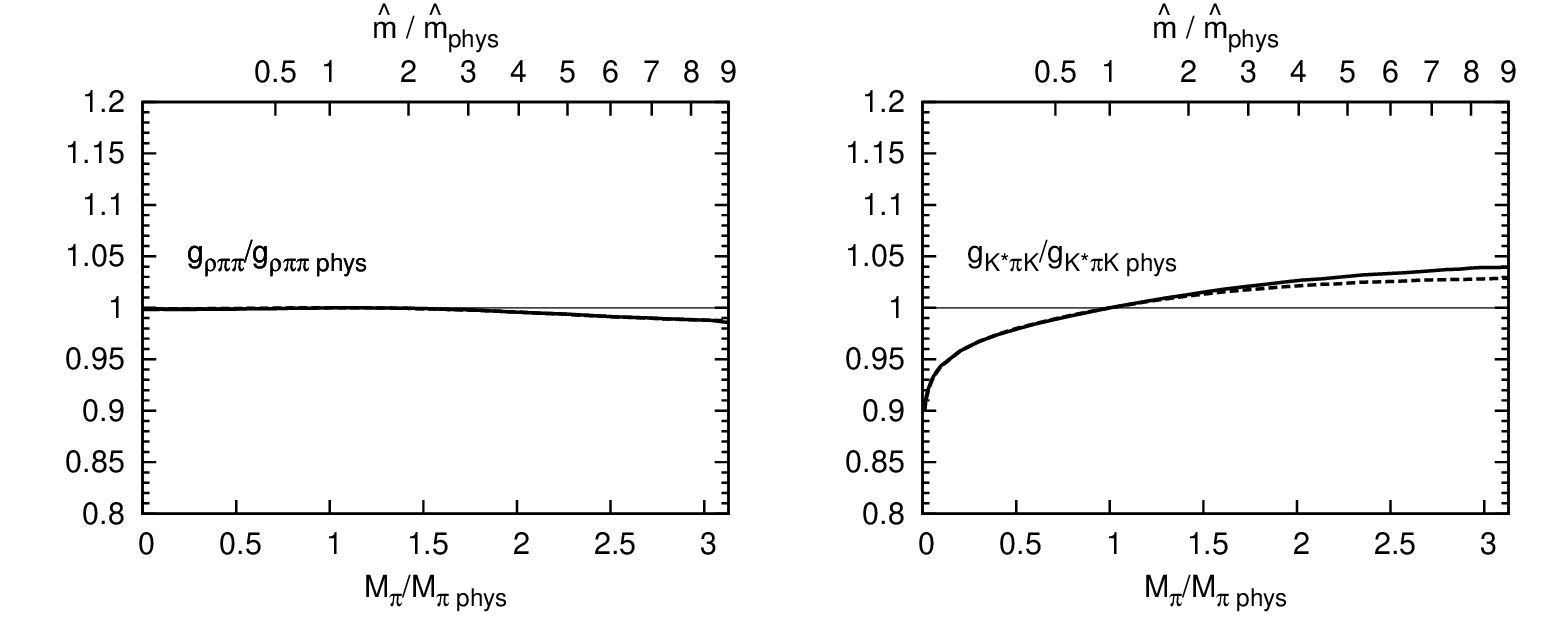}
\parbox{\linewidth} { 
    \caption{ \label{fig:couplingrhoks} 
Two-meson-vector coupling dependence  with respect to
the non-strange quark mass $\hat m$  (horizontal upper scale), or
the pion mass (horizontal lower scale). Note we normalize 
the couplings to their physical values. We show on the left the $\rho(770)$ 
coupling to two pions and on the right that  of the $K^*(892)$ to $\pi K$
(continuous and dashed lines correspond to Fit I and Fit II respectively).
}
}
\end{figure*}

\begin{figure*}
  \centering
  \includegraphics[scale=.72]{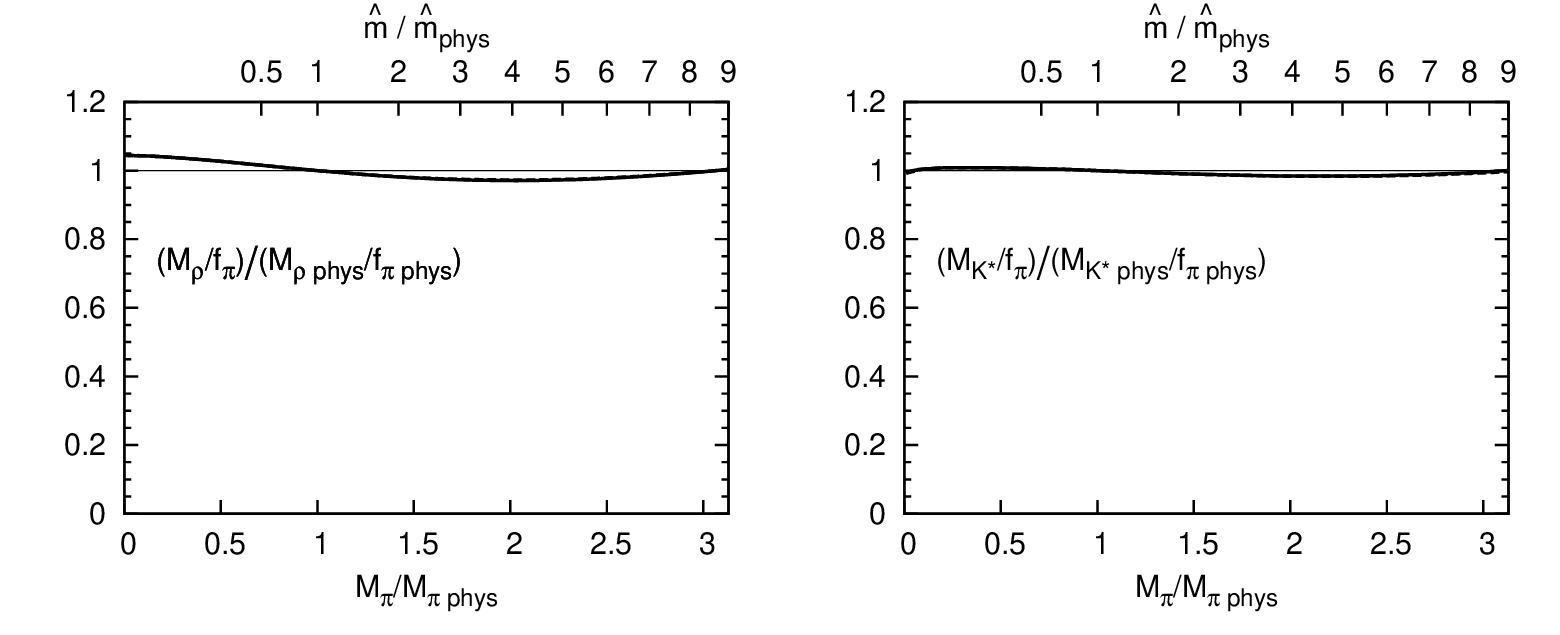}
\parbox{\linewidth} { 
    \caption{ \label{fig:ksfr} 
Ratio of vector resonance masses to the pion decay constant
dependence on the non-strange quark mass (horizontal upper scale), or
the pion mass (horizontal lower scale).  
Once again, we normalize 
all quantities to their physical values.We show the 
$M_\rho/f_\pi$ on the left and $M_{K^*}/f_\pi$ on the right
(continuous and dashed lines correspond to Fit I and Fit II respectively).
Both masses seem to follow the $f_\pi$ quark mass dependence up
to less than 5\% and 2\%, respectively.
}
}
\end{figure*}

\subsubsection{Coupling to two mesons }

The dynamics of resonance-meson-meson interaction is encoded
in the coupling constant that we obtain from the residue of 
the amplitude at the pole position as follows:
\begin{equation}
g^2=-16 \pi \lim_{s \to s_{\rm pole}}
(s-s_{\rm pole})\ t(s)\frac{3}{4\, k^{2}}\label{eq:residue},
\end{equation}
where the normalization factors are chosen 
to recover
the usual expression for the two-meson
width of narrow vector resonances:
\begin{equation}
\Gamma_V=\vert g \vert^2\frac{1}{6\pi}\frac{|{\bf k}|^3}{M_V^2},
\label{eq:V-width}
\end{equation}
$|{\bf k}|$ being the modulus of the meson three-momentum.
Actually, by identifying $\sqrt{s_{\rm pole}}=M_V-i\Gamma_V$, 
we have explicitly checked that we obtain the same numerical
value for the coupling with both equations.
We find $|g_{\rho\pi\pi}|\simeq 6.1$ and $|g_{K^*\pi K}|\simeq 5.5$.

Then, on  Fig.\ref{fig:couplingrhoks} we show the dependence of the
$g_{\rho\pi\pi}$ (left) and the 
$g_{K^*\pi K}$ (right) couplings with respect to the 
pion mass (lower horizontal scale) or 
the non-strange quark mass $\hat m$ (upper horizontal scale).
In order to suppress systematic uncertainties, we
have normalized the couplings to their physical values.
Note that the $g_{\rho\pi\pi}$ is remarkably constant,
deviating from its physical value by 2\% at most, despite the fact that the
quark mass is changed by a factor of 9.
It is also relevant because it justifies
the constancy assumption made in lattice studies
of the $\rho(770)$ width \cite{Aoki:2007rd}.
The $g_{K^*\pi K}$ is also quite independent
of the non-strange quark mass, deviating by 10\% at most in the chiral
limit and by less than 4\% when the quark mass 
is increased by a factor of nine. The results for Fit I and II
are almost indistinguishable.

The constancy of the vector-meson-meson couplings, together with
the classic KSRF relation \cite{KSRF}, provides a striking connection
between the quark mass dependence of the rho mass and the pion decay constant.
Actually, the KSRF relation, obtained from PCAC and vector meson dominance, reads:
\begin{equation}
g_{\rho\pi\pi}^2\simeq M_\rho^2/8f_\pi^2.
\end{equation}
Note that in our calculation we are obtaining $M_\rho$ from a one-loop
ChPT unitarized calculation, whereas $f_\pi$ comes simply from the 
next to leading
order ChPT calculation, but, of course, without unitarization.
It is therefore quite remarkable that the ratio
$M_\rho/f_\pi$ obtained from our amplitudes, shown
in Fig.\ref{fig:ksfr}, is constant within less than 5\% accuracy,
when the quark mass varies by a factor of 9, 
or the pion mass by a factor of 3. 
Note that, as usual, in Fig.\ref{fig:ksfr} we have normalized the ratio
to its physical value.
It seems that the simple KSRF
relation holds remarkably well up to surprisingly 
large values of the non-strange quark mass,
and therefore the $M_\rho$ quark mass dependence 
can be recast with the same factor as that for $f_\pi$.

A similar result is found for the $K^*(892)$
whose ratio $M_{K^*}/f_\pi$ is also shown in Fig.\ref{fig:ksfr}
to deviate by less than 2\% from its physical value.
Note that, according to the second reference in \cite{KSRF},
the $f_K$ dependence does not show up in the relation. 
Actually, had we used 
 $M_{K^*}/\sqrt{f_\pi f_K}$ instead, the deviation 
would have been a factor of 3 larger.

\subsection{Light scalar mesons: the $f_0(600)$ and $\kappa(800)$}
\label{sec:lightscalarmesons}

The $f_0(600)$, or sigma, and the $\kappa(800)$ scalar mesons 
are still somewhat controversial. The main problem is their huge width
that makes their experimental identification complicated.
Despite the fact that their pole mass and width has been
determined by several groups with the help of model independent 
dispersive techniques (with and without ChPT input) and a fairly reasonable
agreement (see \cite{Dobado:1996ps,Caprini:2005zr,Buettiker:2003pp} for recent determinations), 
they are still cited with extremely cautious and
conservative estimates in the PDG \cite{PDG}. Their nature is even more 
controversial, and as commented above, there are no present lattice calculations
with realistic quark masses that could shed some light on the problem.
It is therefore even more interesting to obtain predictions on their 
quark mass dependence. Compared with the vector case, there is an additional 
complication because now we do not necessarily expect a similar
behavior between the $\kappa(800)$ and the $f_0(600)$, since although the 
former should belong to an SU(3) octet, the  latter could be in the singlet, 
the octet, or have a significant mixture of both. As a matter of fact, there are 
indications that its singlet component is actually dominant 
\cite{Black:1998wt,Oller:2003vf}.

\subsubsection{Mass and width}

\begin{figure*}
  \centering
  \includegraphics[scale=.73]{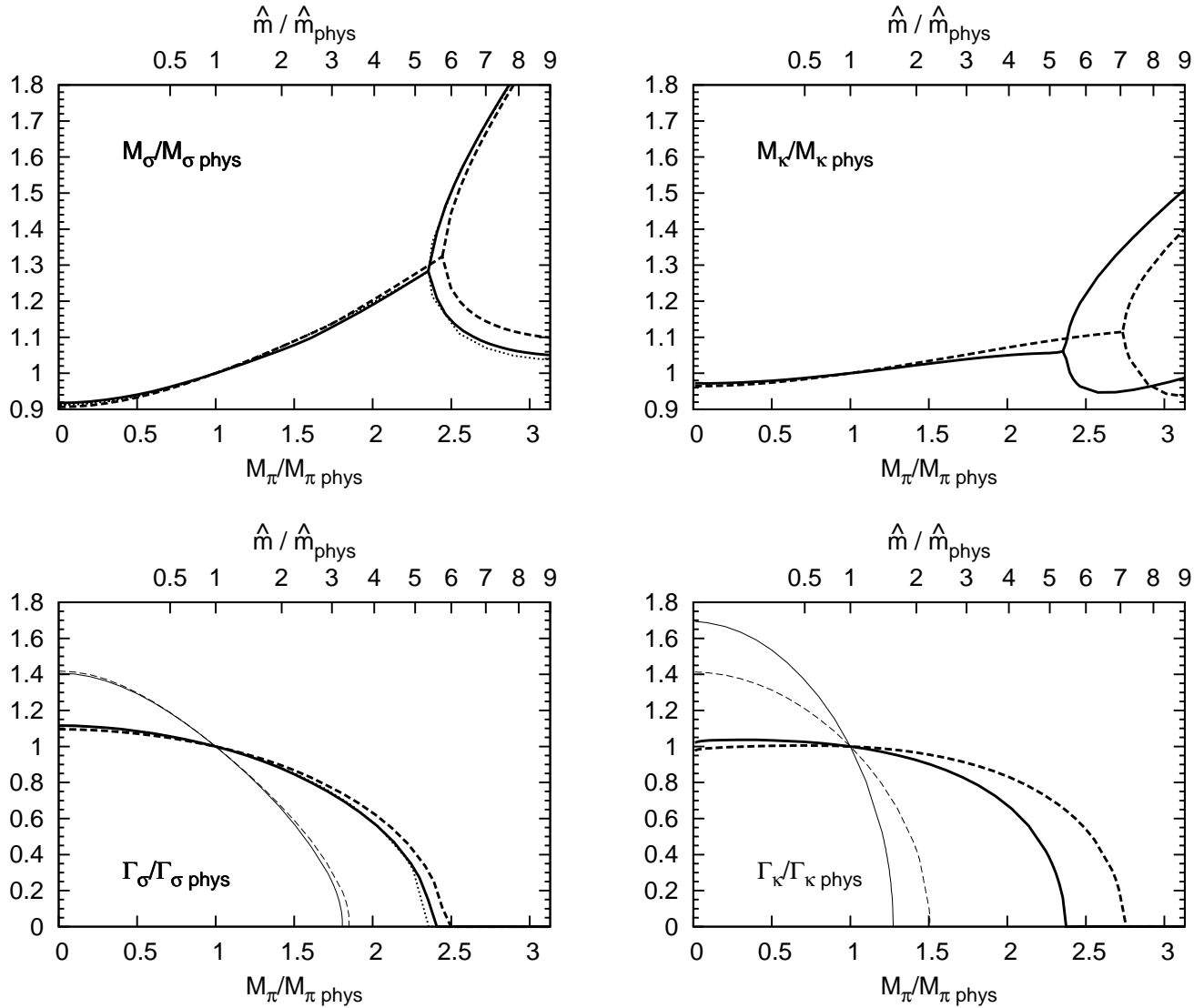}
  \caption{ \label{fig:sigmakappamhatdepend}
Dependence of the $f_0(600)$ 
 and $\kappa(800)$ mass and width with respect to
the non-strange quark mass $\hat m$  (horizontal upper scale), or
the pion mass (horizontal lower scale).
Note that we give all quantities normalized to their 
physical values.  The thick continuous and dashed lines correspond, respectively, to
Fit I and Fit II described in the text with unitarized SU(3) ChPT.
For the $f_0(600)$ these results are very 
compatible with those in \cite{Hanhart:2008mx}
using  SU(2) ChPT (dotted line). 
Let us remark that both resonances, being scalar, develop two poles
on the real axis for sufficiently high masses. The thin lines
show the decrease of the widths if it were only due to phase space reduction 
(the thin continuous line corresponds to Fit I and the dashed one to Fit II). 
}
\end{figure*}
As  the data shows in Fig.~\ref{fig:experiment},
the sigma and kappa resonances do not present a peak nor 
a Breit-Wigner shape in the meson-meson scattering $(I,J)=(0,0)$ and $(1/2,0)$ waves, respectively. 
Once again,
these masses and widths are defined
from the position of their associated pole in the second Riemann sheet,
as follows: $\sqrt{s_{pole}}\equiv M-i\,\Gamma/2$, but one should keep in mind
that these scalar states do not present the typical Breit Wigner shape, so there is no
immediate  equivalence of the mass in terms of a peak 
in the cross section or a time delay
in the propagation.

In Fig.\ref{fig:sigmakappamhatdepend} we show the pole mass and width dependence
of light scalar resonances  on the non-strange quark mass. 
As in Fig.\ref{fig:rhoksmhatdepend}, we show quantities normalized to their physical values and 
we provide two scales
for the  horizontal axis: $\hat m/\hat m_{\rm phys}$ (upper horizontal axis)
and $M_\pi/M_{\pi\, phys}$ (lower horizontal axis). Once again, the continuous line
represents the results for Fit I,  the dashed line those of Fit II, and the dotted line 
stands for the results of unitarized SU(2) ChPT for the $f_0(600)$.
As before we find that the Fit I and II are very consistent with each other, 
and, for the $f_0(600)$ also with the existing SU(2) calculation of 
\cite{Hanhart:2008mx}.

The most prominent feature of the scalars behavior
is the appearance of two branches for the mass as defined above,
already observed for the $\sigma$ in \cite{Hanhart:2008mx}.
The reason is that for physical values of the quark mass,
the  poles associated with resonances appear as
conjugated poles in the second Riemann sheet, i.e., there are poles
at $\sqrt{s_{pole}}\equiv M\pm i\,\Gamma/2$. Of course, only the one in the lowest
half plane is continuous with the physical amplitude in the real axis, and this is 
the one responsible for the physical resonance. 
However, as the quark mass increases these poles move closer to the real axis 
until  they join in a single pole below threshold, but still 
in the second Riemann sheet. If the quark mass is increased further, the poles
split again but without leaving the real axis. The position of each 
one of these poles corresponds to each one of the branches that we show in 
the upper panel of Fig.\ref{fig:sigmakappamhatdepend}.

Although this qualitative
behavior is a well known possibility  for potentials in scalar channels,
one-loop unitarized ChPT is predicting the quark mass value
for which it occurs, which is a genuine prediction for QCD.
For scalar-isoscalar $\pi\pi$ scattering it  was already 
observed in the SU(2) case \cite{Hanhart:2008mx}.
Here we are confirming this position when using SU(3) instead of SU(2) ChPT,
but we see it also happening for the $\kappa(800)$, although the point at which
it happens depends more on the set of LECs. For this reason, we 
think that the existence
of this non-analyticity of the $\kappa(800)$ pole is robust, 
but not so much the precise quark mass value where it occurs.

This ``apparent splitting'' cannot occur for higher partial waves since they
all carry a $k^{2J}$ factor that forces the conjugated poles to join
the real axis exactly at threshold, and then 
one of them jumps to the first Riemann sheet.  

Apart from the evident qualitative similarities between the 
behavior of the $f_0(600)$ 
and the kappa, it is also clear that quantitatively they behave 
somewhat differently.
In particular, the growth of the $\kappa(800)$ mass before the ``splitting point'' is 
much softer than for the $f_0(600)$, 
and even softer than the $\rho(770)$ and $K^*(892)$ growth shown 
in Fig.\ref{fig:rhoksmhatdepend} (please note the difference in scales between 
both figures).

In the lower panels of Fig.\ref{fig:sigmakappamhatdepend}
 we show the quark mass dependence
of the sigma and kappa widths. On the left we show that the 
decrease of the
sigma width we find with the SU(3) one-loop IAM 
is very consistent between Fits I and II, and confirm
the previous results within SU(2) \cite{Hanhart:2008mx}.
On the right we show the results for the $\kappa(800)$ width.
We also show that the width decrease for both of
them cannot be attributed to the phase space reduction, 
due to the increase of pion and kaon masses,
naively expected from the
narrow width approximation
\begin{equation}
\Gamma_S=\vert g \vert^2\frac{1}{8\pi}\frac{|p|}{M_S^2} \label{eq:S-width},
\end{equation}
which we show as a thin continuous (dashed) line corresponding to Fit I (II).
Although the shape of the decrease is slightly different for the $\sigma$ and 
$\kappa$, both scalars behave very differently than vector mesons. Actually, 
we will see next that this implies that the scalar couplings to two mesons 
have a much stronger quark mass dependence than the vector ones.

\subsubsection{Coupling to two mesons }

\begin{figure*}
  \centering
  \includegraphics[scale=.73]{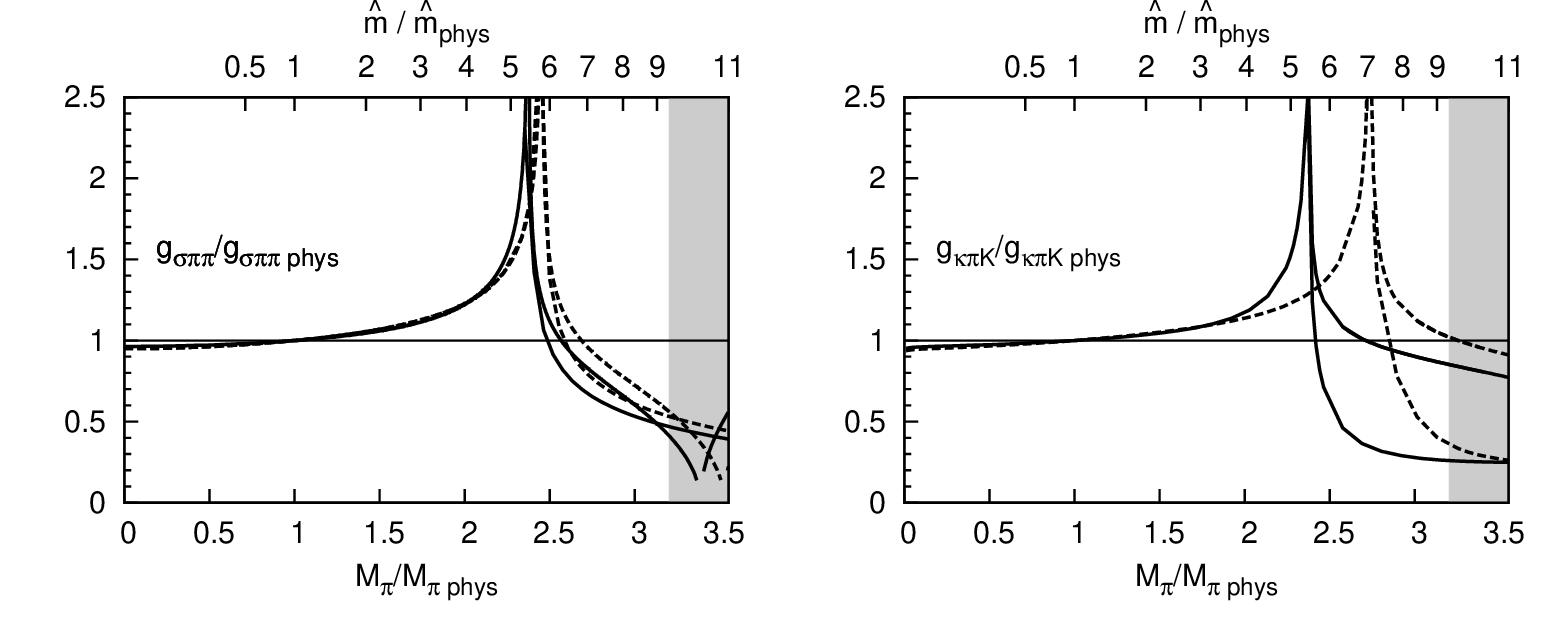}
\parbox{\linewidth} { 
    \caption{ \label{fig:couplingsigmakappa} 
Two-meson-scalar coupling dependence with respect to
the non-strange quark mass $\hat m$  (horizontal upper scale), or
the pion mass (horizontal lower scale). Note we normalize 
the couplings to their physical values. We show on the left the $f_0(600)$
or ``sigma'' 
coupling to two pions and on the right that of the $\kappa(800)$ to $\pi K$.
Let us remark that the two poles in the real axis show different couplings, 
which explains the doubling of the lines above the double branch point.
}
}
\end{figure*}

As we have just seen, the narrow width approximation in Eq.\eqref{eq:S-width}
above is of little use for scalars. But, of course, we can still extract 
the coupling constant
from the residue as we did for vectors, although now the equation reads
\begin{equation}
g^2=-16 \pi \lim_{s \to s_{\rm pole}}
(s-s_{\rm pole})\ t(s)\label{eq:residuescalar},
\end{equation}
We find 
$|g_{\sigma\pi\pi}|\simeq 2.86 \,\gev$ and 
$|g_{\kappa\pi K}|\simeq 3.6 \,\gev$,
to be compared to
$|g_{\sigma\pi\pi}|\simeq 2.97\pm0.04 \,\gev$ and
$|g_{\kappa\pi K}|\simeq 4.94\pm0.07 \,\gev$,
obtained in \cite{Oller:2003vf} or the $|g_{\sigma\pi\pi}|\simeq 2.2$ average obtained in 
\cite{Kaminski:2009qg}.
The agreement is fairly reasonable, taking into account that the data that have been used,
the $\sigma$ and $\kappa$ poles, and the models in those references
differ substantially for each reference.

Thus, in Fig.\ref{fig:couplingsigmakappa} we show 
the quark mass dependence (upper horizontal scale) or pion mass dependence (lower horizontal scale)
of $g_{\sigma\pi\pi}$
and $g_{\kappa\pi K}$. As usual, all quantities are normalized to their physical values.  
Compared with Fig.\ref{fig:couplingrhoks} (note the different scales), we see that
these couplings show a much stronger quark mass dependence. Moreover, they 
increase dramatically near the point of the ``apparent splitting''.
Beyond that point there are two non-conjugate poles lying on the real
axis below threshold in the 
second Riemann sheet.  For this reason, after the splitting point
we plot two curves for each fit. The lowest curve corresponds to the pole 
closest to the threshold, that eventually jumps into the first
Riemann sheet. This threshold crossing from one sheet to the other
corresponds to the point where the coupling tends to zero in the figures,
in good agreement with the well known 
result in \cite{Weinberg:1962hj}. Actually 
this can be checked numerically, because, as shown
in \cite{Gamermann:2009uq} the coupling is inversely proportional to the energy 
derivative of the one-loop function ( $G(s)$ in \cite{Gamermann:2009uq} and $J(s)$ in ChPT \cite{GomezNicola:2001as}),
which is divergent at threshold. Despite this consistency check, within 
our approach this occurs at pion masses close to the naive applicability limit,
and therefore the exact $M_\pi$ value when this happens is not very reliable.

\section{Dependence on the strange quark mass}

Up to here we have only been changing the values of the non-strange quark mass
keeping $m_s$ fixed.
However, since we are 
dealing with the full SU(3) ChPT formalism, we are now able to change
the strange quark, keeping $\hat m$ fixed.  
The dependence of hadronic observables
on the strange quark mass is also of interest for lattice studies and 
for cosmological considerations \cite{Cosmology}.
As we explained in Sect.\ref{sec:ChPT} we will only vary the strange quark mass
in the  limited range $0.7<m_s/m_{s\,\rm phys}<1.3$ to ensure that the kaon does not become
too heavy to spoil the ChPT convergence nor too light to require a coupled channel
formalism to deal with the $K^*(892)$ or $\kappa(800)$ resonances, 
thus introducing additional model dependences in our approach.

\subsection{Light vector mesons: the $\rho(770)$ and $K^*(892)$}

\subsubsection{Mass and width}

\begin{figure*}
  \centering
  \includegraphics[scale=.73]{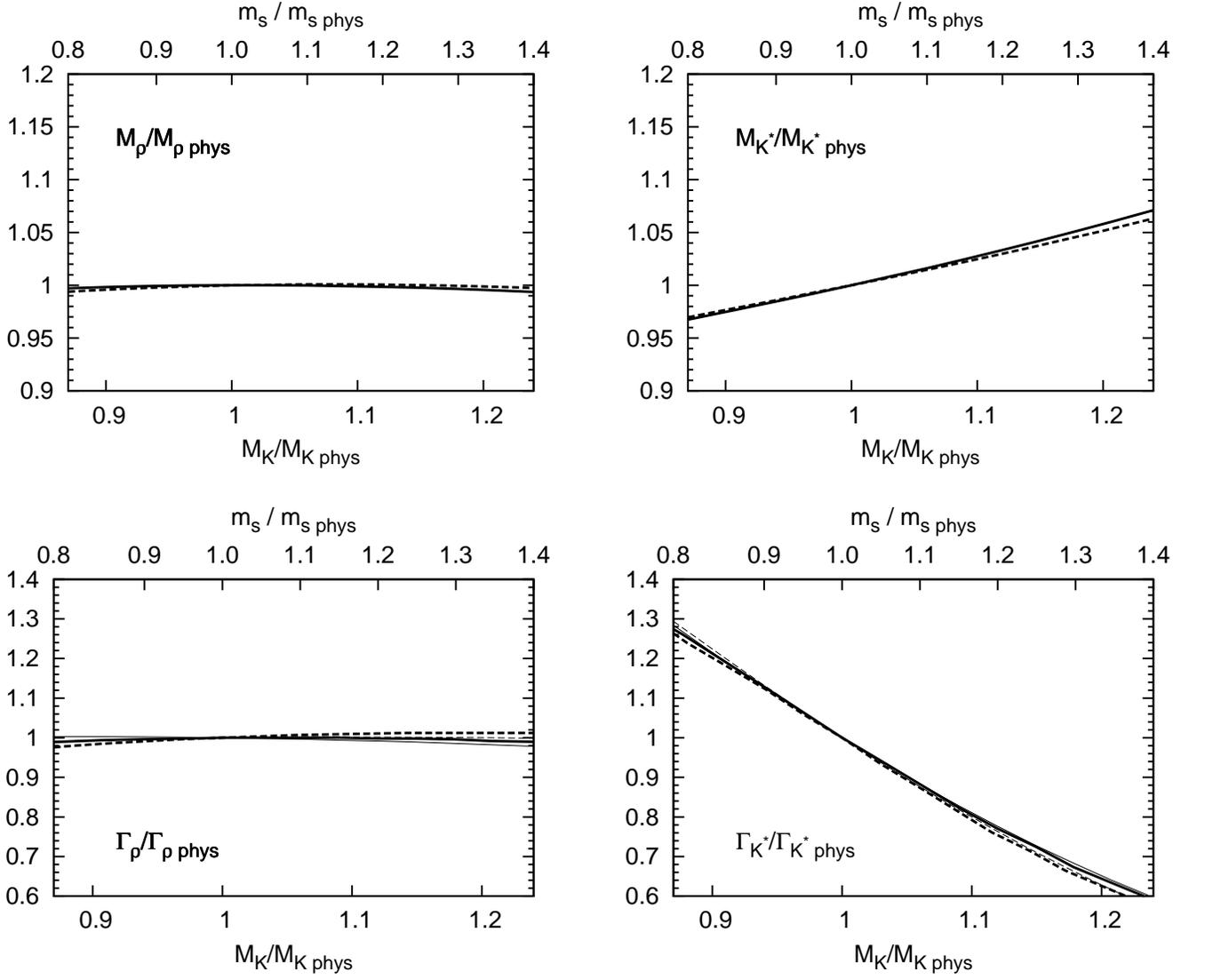}
    \caption{ \label{fig:rhoksmsdepend} 
Dependence of the $\rho(770)$ 
 and $K^*(892)$ mass and width with respect to
the strange quark mass $m_s$  (horizontal upper scale), or
the kaon mass (horizontal lower scale).
Note that we give all quantities normalized to their 
physical values. The thick continuous and dashed lines correspond, 
respectively, to 
Fit I and Fit II described in the text with unitarized SU(3) ChPT.
The thin lines
show the decrease of the widths as if it were only due to phase space reduction 
(the thin continuous line corresponds to Fit I and the dashed one to Fit II).
}
\end{figure*}

As in previous sections we define the mass and 
width of the vector resonances from
the position of their associated poles. Thus, in 
the upper panels of Fig.\ref{fig:rhoksmsdepend} 
we show the quark mass dependence (or kaon
 mass dependence in the lower horizontal scale)
of the $\rho$ and $K^*(892)$ masses.
In the lower panels we show the dependence of their widths.
As usual, all quantities are normalized to their physical values to suppress 
systematic uncertainties.

As it could be expected, both the mass and width of
 the $\rho(770)$, being non-strange, 
are almost independent of the strange quark mass within the range of study. 
Note that the $\rho$ mass actually decreases very slightly, by roughly 1\%. 
Since the pion mass almost remains constant 
-- see Eq.\eqref{pimass} and the $L_6, L_4$ values in Table~\ref{tab:LECs} --, 
this implies that phase space decreases slightly for smaller strange quark mass
and the $\rho(770)$ width decreases accordingly. Actually, we can check 
in  Fig.\ref{fig:rhoksmsdepend} that the width reduction follows
remarkably well the phase space reduction expected from Eq.\eqref{eq:V-width}
(thin continuous and dashed lines).

Looking now at the right panels of Fig.\ref{fig:rhoksmsdepend},
we notice that, as expected,  the $K^* (892)$ shows a much stronger dependence
than the $\rho(770)$
on the strange quark or the kaon masses. 
On the one hand, when the kaon mass is made lighter, the $K^*(892)$ mass decreases,
as it happened when changing the light quark mass, although much faster, i.e., 
up to 5\% when the kaon mass decreases by 20\%.  
Nevertheless,  and contrary to what happened when reducing $\hat{m}$,
the  $K^*(892)$
 width increases significantly, up to 40\%. 
This is due to the fact that the
 $K^*(892)$ decays to $\pi K$, but the kaon mass decrease
is faster than that of the $K^*(892)$.
On the other hand, when the kaon mass is made heavier, the $K^*(892)$ mass grows, but much slower
than the kaon mass, so that phase space shrinks and the resonance width
decreases once more.
We are also showing as thin lines
the expected variation of 
the widths if their only quark mass dependence came
from the change in the particles masses 
and the naive phase space suppression 
in Eq.\eqref{eq:V-width} (thin continuous for Fit I and thin dashed
for Fit II). We see that they are in very good agreement
with our results from the IAM, which suggests that their coupling
to two mesons is almost independent of the quark masses, which we will see next.

\subsubsection{Coupling to two mesons }

\begin{figure*}
  \centering
  \includegraphics[scale=.73]{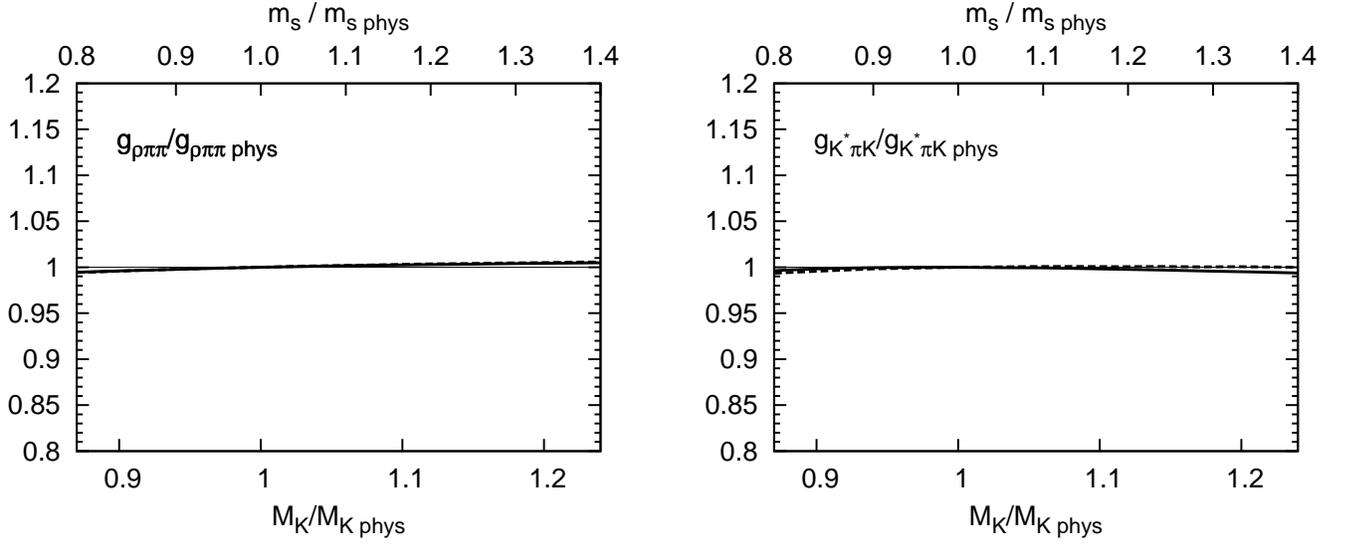}
\parbox{\linewidth} { 
    \caption{ \label{fig:couplingrhoksms} 
Two-meson-vector coupling dependence with respect to
the strange quark mass $m_s$  (horizontal upper scale), or
the kaon mass (horizontal lower scale). Note we normalize 
the couplings to their physical values. We show on the left the $\rho(770)$ 
coupling to two pions and on the right that of the $K^*(892)$ to $\pi K$.
}
}
\end{figure*}

Thus, in Fig.~\ref{fig:couplingrhoksms} we show the dependence 
both on $m_s$ and kaon masses
of the vector to meson-meson couplings. As usual everything is normalized to their physical values.
It can be noted that within the range of variation under study, which is 30\% for
the strange quark mass in either direction, both
the $g_{\rho\pi\pi}$ and $g_{K^*\pi K}$ couplings change by 1\% at most.

\begin{figure*}
  \centering
  \includegraphics[scale=.73]{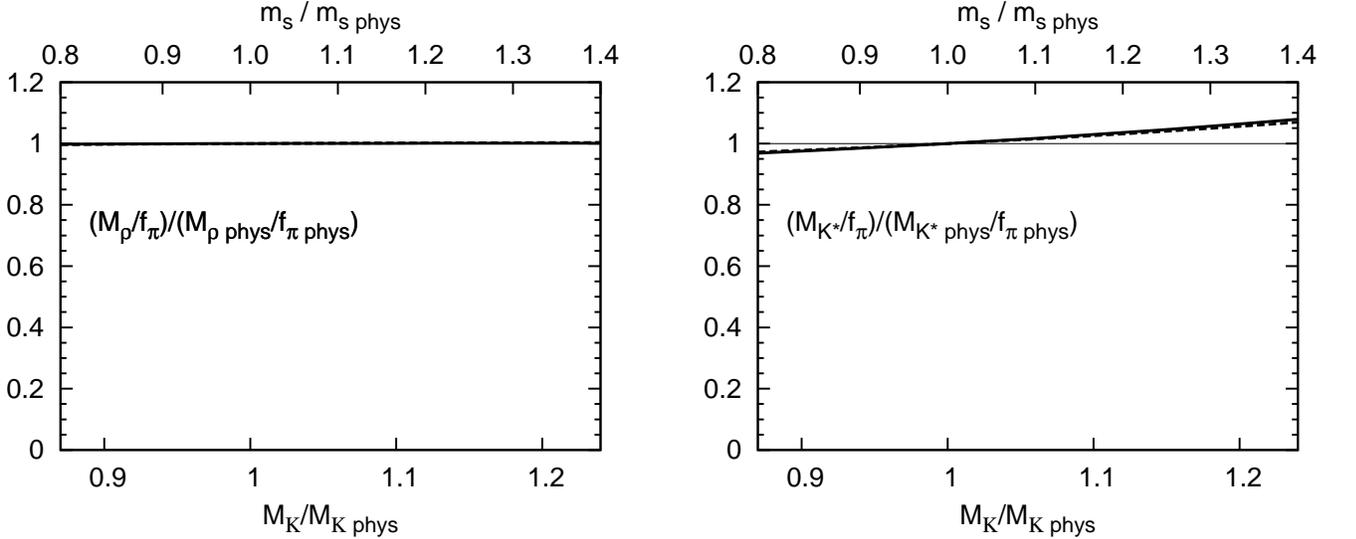}
\parbox{\linewidth} { 
    \caption{ \label{fig:ksfrms} 
Ratio of vector resonance masses to the pion decay constant
dependence on the strange quark mass (horizontal upper scale), or
the kaon mass (horizontal lower scale).  
Once again, we normalize 
all quantities to their physical values.We show the 
$M_\rho/f_\pi$ on the left and $M_{K^*}/f_\pi$ on the right
(continuous and dashed lines correspond to Fit I and Fit II respectively).
For the $\rho$ the result is trivial since we have already shown
that its coupling does not depend on $m_s$. However the $K^*(892)$ mass
deviates from the $f_\pi$ quark mass dependence
by less than 5\% with respect to the physical value.
}
}
\end{figure*}

In Fig.~\ref{fig:ksfrms} we show the results for the KSRF relation
variation in terms of the strange quark mass. Since the $\rho$ coupling has virtually no dependence
on $m_s$, the relation remains trivially constant. For the $K^*(892)$ the relation
is well satisfied (to within less than 5\% from the physical value) in the whole $m_s$ range of our study.

\subsection{Light scalar mesons: the $f_0(600)$ and $\kappa(800)$}

We simply repeat the procedure we used to study the light quark variation 
in Sec. \ref{sec:lightscalarmesons},
but this time changing the strange quark mass instead, and keeping $\hat m$ fixed.

\subsubsection{Mass and width}

\begin{figure*}
  \centering
  \includegraphics[scale=.73]{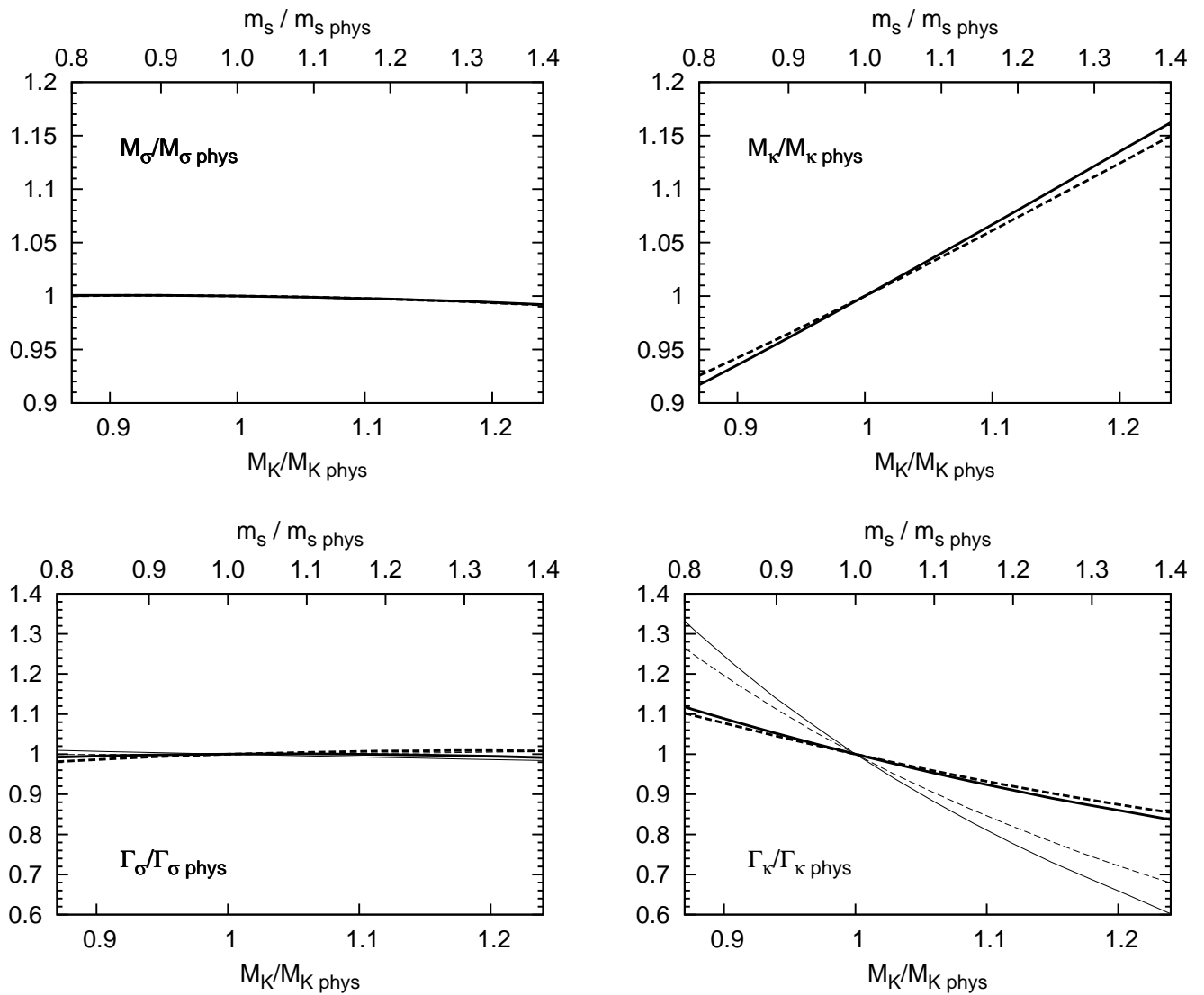}
  \caption{ \label{fig:sigmakappamsdepend}
Dependence of the $f_0(600)$ 
 and $\kappa(800)$ mass and width with respect to
the strange quark mass $m_s$  (horizontal upper scale), or
the kaon mass (horizontal lower scale).
Note that we give all quantities normalized to their 
physical values.  The thick continuous and dashed lines correspond, respectively, to 
Fit I and Fit II described in the text with unitarized SU(3) ChPT.
The thin lines
show the decrease of the widths as if it were only due to phase space reduction 
(the thin continuous line corresponds to Fit I and the dashed one to Fit II)
}
\end{figure*}

Thus, in Fig.~\ref{fig:sigmakappamsdepend}, we show the variation of the sigma 
and $\kappa(800)$ masses and widths with respect to the kaon mass variation 
(lower horizontal scale) or the strange quark mass (upper horizontal scale).
Once again all masses are normalized to their physical values.
As it could be expected, we see in the left panels that the change on the sigma is 
smaller than 1\% on both mass and width
(beware we have changed the scale with respect to 
the previous Fig.~\ref{fig:sigmakappamhatdepend} to make the changes more visible).

A much bigger effect is seen for the $\kappa(800)$ in the right panels, whose mass 
changes by as much as 12\% from its physical value
within the range of study, whereas the width changes by as much as 20\%. However, 
its mass dependence, despite being somewhat stronger than for its vector counterpart 
$K^*(892)$, is still softer than for the kaon itself.
This is the reason why, as the $\kappa(800)$ becomes lighter its width increases, and viceversa.

In the lower panels we have also plotted the expected naive phase space reduction.
This time, however, as the sigma properties barely depend on the strange quark mass,
we only see a significant deviation from that naive behavior in the case of the $\kappa(800)$.

\subsubsection{Coupling to two mesons }

\begin{figure*}
  \centering
  \includegraphics[scale=.73]{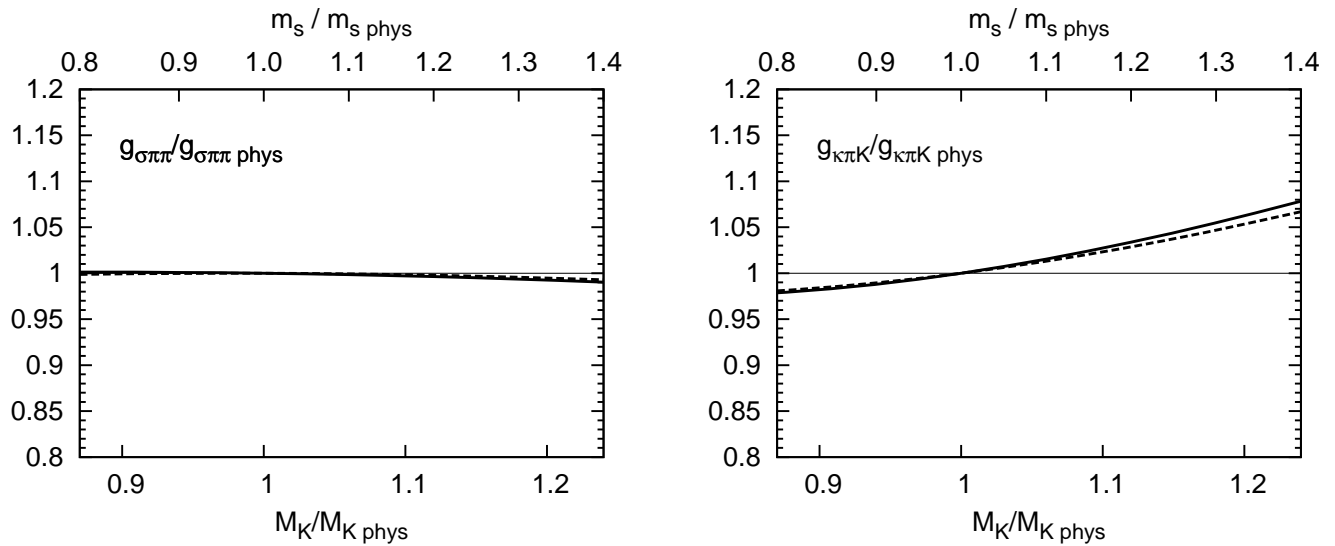}
\parbox{\linewidth} { 
    \caption{ \label{fig:couplingsigmakappams} 
Two-meson-scalar coupling dependence with respect to
the strange quark mass $m_s$  (horizontal upper scale), or
the kaon mass (horizontal lower scale). Note we normalize 
the couplings to their physical values. We show on the left the $f_0(600)$
or ``sigma'' 
coupling to two pions and on the right that of the $\kappa(800)$ to $\pi K$.
}
}
\end{figure*}

For all means and purposes, with respect to strange quark mass variations, the sigma coupling 
to two mesons turns out to be
a constant within our approximation, as can be seen in the left panel of
 Fig.~\ref{fig:couplingsigmakappams}.

In contrast, the $g_{\kappa\pi  K}$ coupling shows some dependence on the strange quark mass.
Actually, it grows by 6\% when the kaon mass is increased by 18\% from its physical value.

\section{Summary and Conclusions}

In this work we have studied the quark mass dependence 
of the light vector and 
scalar resonances generated as poles of meson-meson 
scattering elastic amplitudes
within unitarized one-loop Chiral Perturbation Theory (ChPT). 
This dependence is of interest to relate lattice 
results to hadronic observables, but also
for anthropic and cosmological considerations.
The use of an SU(3) formalism extends previous studies within SU(2), 
allowing us to study the behavior of strange resonances like the 
$\kappa(800)$ and $K^*(892)$, but also to study
variations not only of the light $u$ and $d$ quark masses, but also
of the strange quark mass. 

After a  brief introduction on
how ChPT provides a model independent 
expansion of pion, kaon and eta masses and decay constants,
as well as their two body interaction amplitudes,  
we have reviewed how this series can be used 
inside a dispersion theory formalism  to 
construct the so called Inverse Amplitude Method (IAM)
amplitudes that satisfy elastic unitarity while respecting the ChPT expansion.
It has been known for long that the elastic IAM reproduces well the 
meson-meson elastic scattering data up to 800-1000 MeV, 
including the resonance region. 
Note that we have refrained for the moment to use the
very successful coupled channel IAM precisely because
at present it lacks a dispersive derivation, and we want to
avoid as much model dependence as possible. 
Of course the experimental 
data may fix rather well the energy dependence 
but not so well the mass dependence.
For that reason we have presented  here a new IAM analysis including 
 simultaneously the existing lattice results on meson masses, decay constants and 
scattering lengths. We obtain a fairly good description of experiment
and lattice data using chiral parameters rather similar to existing one and two-loop determinations.
No fine tuning of parameters is required. 
Once this is done, we have varied the quark masses within certain ranges 
that ensure the applicability
of the elastic IAM for the resonances under study:
$\hat m/\hat m_{\rm phys}\leq 9$ and $0.7<m_s/m_{s\,\rm phys}<1.3$ ($\hat m$ is the average mass of the  
$u$ and $d$ quarks). In practice, in ChPT we have changed the squared pion and kaon masses,
which, at leading order, are proportional to quark masses. Although we have shown in Fig.\ref{fig:masses}
that this simple approximation works within roughly 10\% accuracy,
 we have carefully included the full
one-loop corrections, and shown the quark and meson 
mass variation independently in all plots.

In the second Riemann sheet of these amplitudes, the 
IAM generates the  -- conjugated pairs of -- poles associated to
the vector $\rho(770)$, $K^*(892)$ and scalar $f_0(600)$ and $\kappa(800)$ resonances. 
Light vector resonances are well established and there is little relevance
on whether we refer to their ``pole'' 
or Breit-Wigner mass and widths. In contrast,
the scalar $f_0(600)$, or ``sigma'' 
and the $\kappa(800)$ are rather controversial
due to their large apparent width and the lack of a Breit-Wigner shape in the meson-meson
scattering phase shifts. To avoid complications, 
we have always presented our results in terms of ``pole'' definitions of masses,
widths and couplings.

For the $f_0(600)$ and $\rho(770)$ resonances, which appear in $\pi\pi$ scattering,
we have nicely confirmed the similar unitarized one-loop SU(2) ChPT  analysis performed in 
\cite{Hanhart:2008mx}. When increasing $\hat m$
both the sigma and $\rho$ masses grow faster than the pion mass,
whereas their widths decrease. 
However, the $\rho(770)$ mass behaves smoothly in the whole 
quark mass range, whereas, roughly at $M_\pi\sim 340\,$MeV,
 the $f_0(600)$ pole and its conjugated pair meet in the 
second Riemann sheet below threshold, producing a non analyticity --
or ``apparent splitting'' in two branches -- 
of the sigma mass in terms of $M_\pi$.
In addition, we confirm that the $\rho(770)$ 
width decrease, as $\hat m$ grows, follows remarkably well the simple 
expectations of phase space reduction already 
found within the SU(2) formalism. 
Once again, such a simple behavior is not observed
for the sigma.

Of course, the SU(3) formalism allows us now to study also the $K^*(892)$ and
$\kappa(800)$ resonances in $\pi K$ scattering. We find that both the
mass and width of the $K^*(892)$ behave qualitatively and quantitatively 
in a very similar way to those of the $\rho(770)$, which could be expected
given the fact that they belong to the same octet. 
In addition, we have explicitly calculated
here their couplings to two mesons, from the residue
of the partial wave at their associated  pole, finding that they are both
remarkably independent of the non-strange quark mass, 
as suggested from the width behavior.
The $K^*(892)$ coupling is quite well approximated by a constant,
although not so well as in the $\rho$ case.
This could be of relevance when 
computing its width on the lattice as it has already 
been done for the $\rho$ \cite{Aoki:2007rd}.

It therefore seems that light quark masses play 
no significant role in the {\it dynamics} of
the dominant decay modes of vector mesons, 
namely $\rho\rightarrow \pi\pi$ and $K^*\rightarrow \pi K$,
since their couplings seem to be independent
 of light quark masses and all their
width variation can be attributed to the 
phase space modification due to changes in the masses of all
particles.

Furthermore, this provides a hint, checked here by explicit calculation, that
the KSRF relation, that approximates 
these couplings by $ g\simeq M_V/2\sqrt{2} f_\pi$,
holds to less than 5\% when changing $\hat m$ from 0 to 9 times its physical value.
It is remarkable that this relation is so well satisfied, first, because
ours is a one-loop calculation, which, in principle includes higher order pion mass
corrections to KSRF, and the pion mass becomes rather large, 
but, second, because our resonance masses come from unitarized amplitudes
whereas $f_\pi$ stems from the non-unitarized ChPT truncated series.

Concerning the $\kappa(800)$, its behavior is qualitatively similar to that of
the ``sigma'', including the ``apparent mass splitting'' in two branches,
which is a feature that can only occur for scalars. However,
the $\kappa(800)$ non-strange quark mass dependence is softer than for the sigma.
Still the pion mass where the $\kappa(800)$ ``apparent mass splitting'' occurs
is similar to that of the sigma, although with bigger uncertainties
$M_\pi\sim 340-400\,$MeV.
Of course, contrary to the vector case, one could now expect some
differences between the two scalars since they do not necessarily belong to the same octet
and actually, the sigma is believed to be predominantly the singlet state
\cite{Black:1998wt,Oller:2003vf}, and it could even allow for a glueball component.
As we did with the vectors, in this work we have also
calculated explicitly the  behavior of the
scalar couplings to two mesons under quark mass variations.
We find a qualitatively similar behavior for both $g_{\sigma\pi\pi}$ and $g_{\kappa \pi K}$:
contrary to vectors, they cannot be considered constant within the variation
range, particularly when $M_\pi$ comes close to the ``apparent mass splitting'' value,
where it suffers a dramatic enhancement.

Finally, since we use the SU(3) formalism, we have been able
to study the dependence of light resonance properties 
on the strange quark mass. 
Due to the fact that the physical mass of the kaons is already quite high 
but also because we want the $M_K+M_\eta$ threshold to be
significantly above the $K^*(892)$ mass, we have limited
our study to the range $0.7<m_s/m_{s\,\rm phys}<1.3$.
As it could be naively expected, and in contrast to strange resonances,
the masses and widths of 
both the non-strange $\rho$ and $\sigma$ 
are remarkably independent of the strange quark mass.
This time, the $\kappa(800)$ mass has a much stronger dependence 
than that of the $K^* (892)$ -- actually, it grows a factor of three faster.
Once again, the $K^* (892)$ width follows remarkably well 
the behavior dictated by phase space only, and we have checked that its
$\pi K$ coupling is almost independent of $m_s$. The KSRF
relation is also a fairly good approximation in the whole energy range,
 although  not as good as in the case of the non-strange quark.
Concerning the $\kappa(800)$, once again its coupling
is strongly dependent on the quark mass, so that its width does not follow
the naive phase space behavior.

In summary, we have presented an exhaustive study 
on the strange and non-strange quark mass 
dependence of light scalar and vector resonances appearing in elastic
Goldstone bosons scattering.
For the future, this work could be extended to other light scalar 
mesons like the $f_0(980)$ and
$a_0(980)$  using a coupled channel formalism,
that is somewhat less rigorous as it has no dispersive derivation,
and also is much more complicated due to the presence
 of the $K\bar K$ threshold.

To conclude, 
and apart from the interest for studies of constraints on hadronic 
properties from cosmological or anthropic considerations,
we think that the quark mass dependence studied here
will be within the reach of lattice 
studies in the not too distant future 
-- it is already so for the $\rho$ meson --
and we expect our results to be useful in the chiral extrapolation
of lattice results to physical values.

\section*{Acknowledgments}
We thank C. Hanhart, E. Oset and G. R\'{\i}os
 for useful discussions, J.A. Oller for suggesting us
to include the KSRF relation in our study and W. Dunwoodie for 
providing us with lists of experimental $K \pi$ data.
Work partially supported by Spanish Ministerio de 
Educaci\'on y Ciencia research contracts: FPA2007-29115-E,
FPA2008-00592 and FIS2006-03438, 
U.Complutense/Banco Santander grant PR34/07-15875-BSCH and
UCM-BSCH GR58/08 910309. We acknowledge the support 
of the European Community-Research Infrastructure
Integrating Activity
“Study of Strongly Interacting Matter” 
(acronym HadronPhysics2, Grant Agreement
n. 227431)
under the Seventh Framework Programme of EU.

\end{document}